\documentclass[onecolumn, journal]{IEEEtran}

\ifCLASSINFOpdf

\else

\fi

\usepackage{cite}
\usepackage{algorithm}
\usepackage{algcompatible}
\usepackage{amsmath,amssymb}
\usepackage{float}
\usepackage[acronym]{glossaries}
\usepackage[hidelinks]{hyperref}
\usepackage[font=small]{caption}
\usepackage{subcaption}
\usepackage{tkz-graph}
\usepackage{setspace}
\usepackage{stackengine}
\usepackage{amsthm}
\usepackage{color}

\newtheorem{proposition}{Proposition}
\newtheorem{theorem}{Theorem}
\newtheorem*{theorem*}{Wormald's Theorem}
\newtheorem{corollary}{Corollary}
\theoremstyle{definition}
\newtheorem{definition}{Definition}
\theoremstyle{remark}
\newtheorem{remark}{Remark}
\newtheorem{notation}{Notation}
\DeclareRobustCommand{\stirling}{\genfrac\{\}{0pt}{}}

\def\delequal{\mathrel{\ensurestackMath{\stackon[1pt]{=}{\scriptstyle\Delta}}}}

\def\ve#1{{\mathchoice{\mbox{\boldmath$\displaystyle #1$}}%
{\mbox{\boldmath$\textstyle #1$}}%
{\mbox{\boldmath$\scriptstyle #1$}}%
{\mbox{\boldmath$\scriptscriptstyle #1$}}}}

\newacronym{cfc}{\textsc CFC}{Coded \textit{File} Caching}
\newacronym{csc}{\textsc CSC}{Coded \textit{Subfile} Caching}
\newacronym{cfcm}{\textsc CFC-MD}{\acrshort{cfc} with online Matching Delivery}
\newacronym{cfcc}{\textsc CFC-CCD}{\acrshort{cfc} with  Greedy Clique Cover Delivery}
\newacronym{cscm}{\textsc CSC-MD}{\acrshort{csc} with online Matching Delivery}
\newacronym{cscc}{\textsc CSC-CCD}{\acrshort{csc} with online Clique Cover Delivery}

\begin{document}
\title{Decentralized Coded Caching Without File Splitting}

\author{Seyed Ali Saberali, \IEEEmembership{˜Member,˜ IEEE}, Lutz Lampe, \IEEEmembership{˜Senior  Member,˜ IEEE},\\ and Ian Blake, \IEEEmembership{˜Fellow,˜ IEEE}
}

\maketitle
\begin{abstract}
Coded caching is an effective technique to reduce the redundant traffic in wireless networks. The existing coded caching schemes require the splitting of files into a possibly large number of subfiles, i.e., they perform coded subfile caching.
Keeping the files intact during the caching process would actually be appealing, broadly speaking because of its simpler implementation. However, little is known about the effectiveness of this coded file caching in reducing the data delivery rate.
In this paper, we propose such a file caching scheme which uses a decentralized algorithm for content placement and  either a greedy clique cover or an online matching algorithm for the delivery of missing data. We derive approximations to the expected delivery rates of both schemes using the differential equations method, and show them to be tight through concentration analysis and computer simulations.
Our numerical results demonstrate that the proposed coded file caching is significantly more effective than uncoded caching in reducing the delivery rate.
We furthermore show the additional improvement in the performance of the proposed scheme when its application is extended to subfile caching with a small number of subfiles.
\end{abstract}

\begin{IEEEkeywords}
5G communications, clique cover algorithm, coded file caching, index coding,  traffic offloading.
\end{IEEEkeywords}

\section{Introduction}\label{sec:intro}
Caching of popular content at the wireless edge is a promising technique to offload redundant traffic from the backhaul communication links of the next generation wireless networks \cite{Liu:2016,Zeydan:2016,Debbah:2016,Tulino:2017,Golrezaei:2013}. An integral part of the 5G cellular systems is the dense deployment of small-cells within macrocells to increase the spectral efficiency \cite{Golrezaei:2013,Qiao:2016,Han:2016}. 
By having each small base-station equipped with a large memory storage, multiple caching nodes co-exist within each macrocell. This provides the opportunity to use network coding to further decrease the backhaul traffic of the macrocell over the conventional caching systems \cite{Maddah_Magazine:2016}. In particular, the cached content can be used as side information to decode  multicast messages that simultaneously deliver the missing content to multiple caches. The design of content placement in the caches and construction of the corresponding coded multicast messages are two elements that constitute the coded caching problem \cite{Maddah_Magazine:2016,Maddah_limits:2014,Maddah_decentralized:2014}. 

\subsection{Preliminaries}
Coded caching is closely related to the index coding with side information problem \cite{Aalon:2008,Bar-Yossef:2011}. In both cases,  there is a server that transmits data to a set of $K$ caching clients over a broadcast channel. The server is aware of the clients' cached content. Each client wants certain blocks of data, yet some of these blocks might be missing from its cache.  The objective is to transmit the minimum amount of supplemental data over the broadcast channel such that all clients can derive the data blocks that they requested \cite{Bar-Yossef:2011,Maddah_limits:2014}. In the coded caching literature, the amount of supplemental data transmitted is referred to as delivery rate \cite{Maddah_limits:2014}. The main factor that differentiates the two problems is that in index coding, the cached content is usually given and the focus is on the design of the server messages. However, in coded caching, the placement of the content in the caches can also be designed.  

An information theoretic formulation of coded caching was developed in \cite{Maddah_limits:2014}. The authors proposed a coded caching scheme  which uses a centralized content placement algorithm and a set of coded delivery messages. The worst-case delivery rate of this scheme was shown to be  $RF=\frac{K(1-M/N)}{1+KM/N}F$ packets, where $M$ and $N$ are the number of files that each cache can store and the total number of files that are predicted to be popular, respectively. Parameter $F$ is the number of packets per file. A packet can be a single bit or a chunk of bits of a file, but they are all of the same length and cannot be broken into smaller parts.  Quantity $R$ is the delivery rate in equivalent number of files. The placement of \cite{Maddah_limits:2014} splits each file into a fixed number of subfiles. A subfile is a set of packets of a file. We refer to any coded caching that breaks files into subfiles as \acrfull{csc}. Notice that ${1\!+\!KM/N}$ in the rate expression is the multiplicative gain due to coding.
In \cite{Maddah_decentralized:2014}, the authors  proposed a decentralized \acrshort{csc} which allows every cache to store the content independent from the content of the other caches. The proposed scheme preserves most of the coding gain of the centralized scheme of \cite{Maddah_limits:2014} and has a worst-case delivery rate of $RF\!=\!(N/M\!-\!1)(1\!-\!(1\!-\!M/N)^K)F$ packets in the asymptotic regime of $F\rightarrow+\infty$. Since decentralized caching does not require any central coordination among the caches for placement, it is the preferred caching framework for the next generation wireless systems. As a result, the scheme of \cite{Maddah_decentralized:2014} has served as the building block of several other coded caching methods \cite{Maddah_nonuniform:2014,Maddah_online1:2014,Maddah_hier:2014,Digavi2:2015,Zhang:2015}. Both \cite{Maddah_limits:2014} and \cite{Maddah_decentralized:2014} considered the worst-case delivery rates which correspond to the demand vectors where all caches' requests are distinct. In a recent work \cite{Maddah_optimal:2016}, the minimum \textit{average} delivery rates of both centralized and decentralized coded cachings with uncoded prefetching were characterized.

\subsection{Motivation and Related Works}
In this paper, we consider the problem of decentralized \acrfull{cfc}, where files are kept intact and are not broken into subfiles.
Our first motivation to analyze \acrshort{cfc} is the large size requirement of most of the existing decentralized subfile cachings. 
In particular, \cite{Shanmugam:2016} has developed a finite-length analysis of the decentralized coded caching of \cite{Maddah_decentralized:2014} and showed that it has a multiplicative coding  gain of at most 2 even when $F$ is exponential in the asymptotic gain $KM/N$. In \cite{Shanmugam:2016}, the authors have also proposed a new \acrshort{csc} scheme that requires a file size of $\Theta(\lceil N/M \rceil^{g+1}(\log(N/M))^{g+2}(2e)^g)$ to achieve a rate of $4K/(3(g+1))$.

Another important limitation of the existing coded caching schemes is the large number of subfiles required to obtain the theoretically promised delivery rates. For instance, \cite{Maddah_decentralized:2014} requires $2^K$ subfiles per file, and the different missing subfiles of a file requested by each cache are embedded into $2^{K-1}$ different server messages \cite{Maddah_decentralized:2014}. This exponential growth in $K$ has adverse consequences on the practical implementation of the system and its performance. In particular, a larger number of subfiles imposes large storage and communication overheads on the network. As is pointed out in \cite{Tulino:2017}, during placement, the caches must also store the identity of the subfiles and the packets that are included in each subfile, as a result of the randomized placement algorithms used. This leads to a storage overhead. Similarly, during delivery, the server needs to transmit the identities of the subfiles that are embedded in each message which leads to communications overhead. Neither of these overheads are accounted for in the existing coded caching rate expressions, while they are non-negligible when the number of subfiles is relatively large. 
For centralized coded caching, \cite{Shanmugam:2017} has proposed a scheme that requires a linear number of subfiles in $K$ and yields a near-optimal delivery rate. As is pointed out in \cite{Shanmugam:2017}, the construction of the proposed scheme is not directly applicable to the real caching systems as it requires $K$ to be quite large. However, it is interesting to see that a small number of subfiles can in theory result in a near-optimal performance.  For decentralized caching, reference \cite{Tulino:2015} has investigated the subfile caching problem in the finite packet per file regime and has proposed a greedy randomized algorithm, GRASP, to preserve the coding gain in this regime. Based on computer simulations, the authors show that with a relatively small number of subfiles, a considerable portion of the asymptotic coding gain of infinite file size regime can be recovered. This observation further motivates a closer look at the file caching problem and its performance analysis.
Finally, since the  conventional uncoded caching systems do not break files into large numbers of subfiles, it is easier for the existing caching systems to deploy coded caching schemes that require one or a small number of subfiles per file.

\subsection{Contributions}\label{subsec:cont}
Although \acrshort{csc} has been well investigated, the analysis of \acrshort{cfc} is missing from the literature.  Based on the motivations we discussed earlier, it is worthwhile to explore \acrshort{cfc} for its potential in reducing the delivery rate. In this paper, we investigate the decentralized \acrshort{cfc} problem by proposing new placement and delivery algorithms. The placement algorithm is decentralized and does not require any coordination among the caches. The first proposed delivery algorithm is in essence a greedy clique cover procedure which is applied to certain side information graphs. This algorithm is designed for the general settings of \acrshort{cfc}. The second delivery algorithm adapts the online matching procedure of \cite{Mastin:2016} and is suitable for \acrshort{cfc} in the small cache size regime.

The online matching procedure can be seen as a special case of the message construction algorithm in \cite{Maddah_Delay:2015} with the merging procedure limited to 1-level merging.

The main contribution of this paper is the analysis of the expected delivery rates of the proposed algorithms for file caching. In particular, we provide a modeling of the dynamics of these algorithms through systems of differential equations. The resulting system can be solved to provide a tight approximation to the expected delivery rate of  \acrshort{cfc} with the proposed algorithms. We provide a concentration analysis for the derived approximations and further demonstrate their tightness by computer simulations. 

Our results show that \acrshort{cfc} is significantly more effective than uncoded caching in reducing the delivery rate, despite its simple implementation and structurally constrained design. It is shown that the proposed \acrshort{cfc} schemes preserve a considerable portion of the additive coding gain of the state-of-the-art \acrshort{csc} schemes.
We present a discussion on the extension of the proposed placement and delivery algorithms to subfile caching with an arbitrary number of subfiles per file. We show that a considerable portion of the additive gain of the existing \acrshort{csc} methods can be revived by using a small number of subfiles per file, which is consistent with the observations made in \cite{Tulino:2015}.  Hence, our  proposed method provides a means to tradeoff the delivery rate with the complexity caused by the use of a larger number of subfiles.

The remainder of this paper is organized as follows. We present our proposed \acrshort{cfc} method in Section~\ref{sec:ove}. Statistical analysis of the delivery rates and concentration results are provided in Section~\ref{sec:ana}. We present numerical examples and simulation results in Section~\ref{sec:sim}. In Section~\ref{sec:dis}, we discuss the generalization of the proposed placement and delivery to \acrshort{csc}. We conclude the paper in Section~\ref{sec:con}.

\section{Placement and Delivery Algorithms}\label{sec:ove}
\subsection{Problem Setup for \acrshort{cfc}}
Similar to \cite{Maddah_decentralized:2014}, we consider a network with a central server and $K$ caches. The central server is connected to  caches through an error-free broadcast channel. A library of  $N$  popular files is given, where the popularity distribution is uniform over the files. Every file is of the same length of $F$ packets. The storage capacity of every cache is $M$ files.  
For the \acrshort{cfc} problem which we consider here, each cache can only store the entirety of a file and partial storage of a file is not allowed during content placement. This is the main element that differentiates the problem setups of \acrshort{cfc} and \acrshort{csc}.
\begin{figure}[t]
\centering
\includegraphics[width=.355\linewidth]{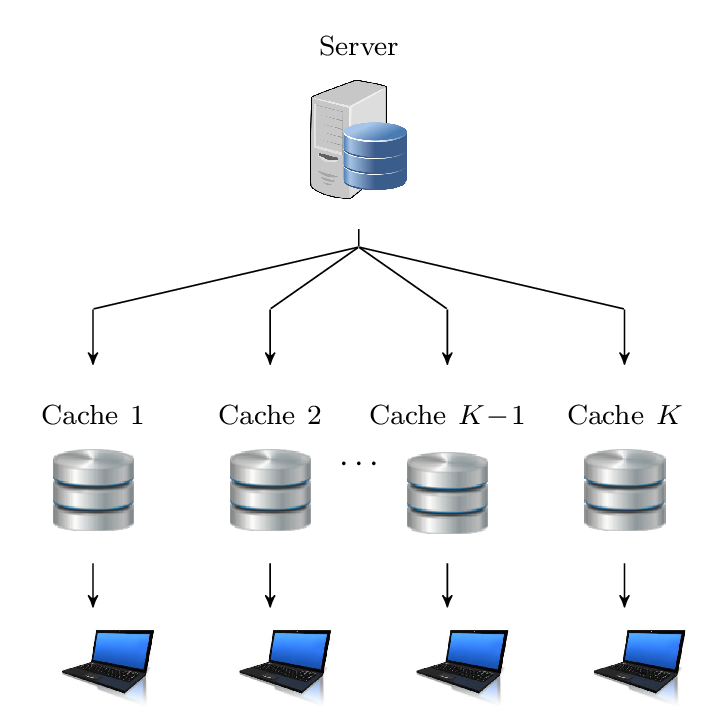}
\caption{Schematic of a network of caches and a central server.}\label{fig:network}
\end{figure}

The delivery phase takes place after the placement, where at each time instant, every cache $k$ reveals one request for a file $d_k$ based on the request of its local users.  We define the demand vector $\ve{d}= [d_1,\ldots,d_K]$ as the vector that consists of the requests made by all the caches at the current time instant. User requests are assumed to be random and independent.  If the requested file is available in the corresponding cache, the user is served locally. Otherwise, the request is forwarded to the server.  The server is informed of the forwarded requests and transmits a signal of size $R(\ve{d};\mathcal{P},A_D)$ files over the broadcast link. The quantity $R(\ve{d};\mathcal{P},A_D)$ is the delivery rate for the demand vector $\ve{d}$, given a specific placement of files $\mathcal{P}$  and a delivery algorithm $A_D$. Placement $\mathcal{P}$ is fixed for all the demand vectors that arrive during the delivery phase. Every cache that has forwarded its request to the server must be able to decode the broadcasted signal for the file it requested. We are interested in minimizing $\mathbb{E}\left(R(\ve{d};\mathcal{P},A_D)\right)$, where the expectation is over the randomness in vector $\ve{d}$, and possible randomness in the delivery algorithm $A_D$ and in the placement algorithm that determines $\mathcal{P}$.

In the following, we propose algorithms for the placement and delivery of \acrshort{cfc}.

\subsection{Placement}
Algorithm~\ref{alg:file_plc} shows the randomized procedure that we propose for content placement. Here, each cache stores $M$ files from the library uniformly at random. The placement is decentralized as it does not require any central coordination among the caches. In Algorithm~\ref{alg:file_plc}, all packets of $M$ files are entirely cached. Notice that each cache stores any particular file with probability $q=M/N$, independent from the other caches. However, the storage of the different files in a given cache are statistically dependent, as the total number of cached files must not exceed $M$.
\begin{algorithm}[t]     
\caption{Decentralized File Placement}          
\label{alg:file_plc}    
\begin{algorithmic}[1]                  
\REQUIRE Library of $N$ files
\FOR{$k=1,\ldots,K$}
\STATE Cache $k$ stores $M$ out of $N$ files from the   
\STATEx \quad  library uniformly at random
\ENDFOR
\end{algorithmic}
\end{algorithm}
Notice that Algorithm~\ref{alg:file_plc} is different from the decentralized subfile placement of \cite{Maddah_decentralized:2014}, as in the latter, an $M/N$-th portion of the packets of \textit{every} file in the library is cached and the packets of each file are cached independently from the packets of the other files.

\subsection{Delivery}\label{subsec:del}
For the proposed delivery, the server forms a certain side information graph for every demand vector that arrives, and delivers the requests by applying a message construction procedure to that graph. The graphs that we use are defined in the following.

\begin{definition}\label{def:digraph}
For a given placement and demand vector, the side information digraph $\mathcal{D}$ is a directed graph on $K$ vertices. Each vertex corresponds to a cache.
A vertex has a loop if and only if the file requested by the cache is available in its storage.
There is a directed edge from $v$ to $w$ if and only if the file requested by cache $w$ is in cache $v$.
\end{definition}
Digraph $\mathcal{D}$ was first introduced in the index coding literature and was used to characterize the minimum length of linear index codes \cite{Bar-Yossef:2011}. This was done through computing the \textit{minrank} of $\mathcal{D}$ \cite[Theorem~1]{Bar-Yossef:2011}, which is an NP-hard problem. 

\begin{definition}\label{def:graph}
For a given placement and demand vector, the side information graph $\mathcal{G}$ is defined on the same set of vertices as $\mathcal{D}$.
A vertex in $\mathcal{G}$ has a loop if and only if it has a loop in $\mathcal{D}$. There is an (undirected) edge between $v$ and $w$ if and only if both edges from $v$ to $w$ and from $w$ to $v$ are present in $\mathcal{D}$.
\end{definition}

We propose two algorithms that make use of graph $\mathcal{G}$ for delivery.
\paragraph{Greedy Clique Cover Delivery Algorithm} For delivery, the server constructs multicast messages with the property that each cache can derive its desired file from such messages using the side information that it has available. 

Consider a set of unlooped vertices in graph $\mathcal{G}$ that form a clique.\footnote{For an undirected simple graph (no loops and no multiple edges), a clique is a subset of vertices where every pair of vertices are adjacent.} Each vertex in that set has the file requested by the other caches available, but its own requested file is missing from its local storage. As a result, if the server transmits a message that is the XOR of the files requested by the caches in that clique, the message is decodeable by all such caches for their desired files \cite[Section~II.A]{Shanmugam:2016}. Hence, to deliver the requests of every cache and minimize the delivery rate, it is of interest to cover the set of unlooped vertices of $\mathcal{G}$ with the minimum possible number of cliques and send one server message per clique to complete the delivery.   Notice that the looped vertices of $\mathcal{G}$ do not need to be covered as the files requested by such caches are available to them locally. 

Finding the minimum clique cover is NP-hard \cite{Karp:1972}. 
In this section, we adapt a polynomial time greedy clique cover procedure for the construction of the delivery messages of \acrshort{cfc}. The proposed clique cover procedure is presented in Algorithm~\ref{alg:cc}.  
\begin{notation}
In Algorithm~\ref{alg:cc}, the content of the file requested by cache $k$ is denoted by $X_k$ and $\oplus$ represents the bitwise XOR operation. 
\end{notation}

\begin{algorithm}[t]     
\caption{Greedy Clique Cover Delivery}          
\label{alg:cc}          
\begin{algorithmic}[1]            
\REQUIRE $\mathcal{G}$, vertex labels $v_1,\cdots,v_K$
\STATEx // Form Cliques
\STATE $C\leftarrow \emptyset$
// initialize set of cliques
\FOR{$t=1,\ldots,K$}
\IF{$v_t$ has a loop}
\STATE Do nothing
\ELSIF{there is a suitable clique for $v_t$ in $C$}
\STATE Join $v_t$ to (one of) the largest suitable clique(s) (randomly) and update $C$ \ELSE
\STATE Add $\{v_t\}$ to $C$ as a clique of size 1
\ENDIF
\ENDFOR
\STATEx // Transmission of Messages
\FOR{$c\in C$}
\STATE Transmit $\oplus_{k\in c}X_k$
\ENDFOR
\end{algorithmic}
\end{algorithm}

In Algorithm~\ref{alg:cc}, a vertex $v_t$ arrives at iteration (time) $t\in\{1,\cdots,K\}$. If $v_t$ is looped, we proceed to the next iteration. Otherwise, we check if there is any \textit{previously formed} clique of size $s\in\{1,\ldots,t-1\}$ that together with $v_t$ forms a clique of size $s+1$. We call such a clique a \textit{suitable clique} for $v_t$. If suitable cliques for $v_t$ exist, we add $v_t$ to the largest one of them to form a new clique. The rationale for choosing the largest suitable clique is explained in Appendix~\ref{app:lar}. If there are multiple suitable cliques with the largest size, we pick one of them uniformly at random.  If no suitable clique exists, $v_t$ forms a clique of size 1. Notice that the cliques formed by Algorithm~\ref{alg:cc} are disjoint and cover the set of unlooped vertices of $\mathcal{G}$. After the clique cover procedure completes, the server sends a coded message corresponding to each clique. 

The use of Algorithms~\ref{alg:file_plc} and \ref{alg:cc} for placement and delivery leads to a caching scheme that we call \acrlong{cfcc} (\acrshort{cfcc}).

\begin{remark}
In Algorithm~\ref{alg:cc}, each cache only needs to listen to one server message to decode the file it requested, as the entirety of the file is embedded in only one message. This is in contrast to the \acrshort{csc} of \cite{Maddah_decentralized:2014} which requires each cache to listen to $2^{K-1}$ out of the $2^K$ server messages  to derive its requested content. 
\end{remark}
Algorithm~\ref{alg:cc} has a worst-case complexity of $O(K^2)$. This is because there are at most $K$ unlooped vertices in $\mathcal{G}$, and in each iteration of the algorithm, the adjacency of $v_t$ with at most $t-1$ vertices must be checked for finding the largest suitable clique. 
\paragraph{Online Matching Delivery Algorithm}
\begin{algorithm}[t]     
\caption{Online Matching Delivery}          
\label{alg:m}          
\begin{algorithmic}[1]                  
\REQUIRE $\mathcal{G}$, vertex labels $v_1,\cdots,v_K$
\STATEx // Transmission of coded messages
\STATE $Q\leftarrow \emptyset$ // set of matched or looped vertices
\FOR{$t=1,\ldots,K$}
\STATE $\mathcal{G}_t\leftarrow$ subgraph induced by $\mathcal{G}$ on vertices
\STATEx  $\quad\qquad\{v_1,\cdots,v_t\}\backslash Q$
\IF{$v_t$ has a loop}
\STATE $Q\leftarrow Q\cup \{v_t\}$

\ELSIF{$v_t$ has a neighbor in $\mathcal{G}_t$}
\STATE  Match $v_t$ to a random neighbor  $v_s\in\mathcal{G}_t$
\STATE Transmit $X_{v_t}\oplus X_{v_s}$
\STATE $Q\leftarrow Q\cup \{v_t,v_s\}$
\ENDIF
\ENDFOR
\STATEx // Transmission of Uncoded Messages
\FOR{$v\in \{v_1,\cdots,v_K\}\backslash Q$}
\STATE Transmit $X_v$
\ENDFOR
\end{algorithmic}
\end{algorithm}
We now propose our second delivery algorithm, which is applicable to the small cache size regime, i.e., when $q$ is small. In this regime, the probability of having large cliques is small. Hence, one can restrict the size of the cliques in the clique cover procedure to reduce the complexity without considerably affecting the delivery rate. As an extreme case, Algorithm~\ref{alg:m} shows an online greedy matching algorithm adapted from \cite[Algorithm~10]{Mastin:2016}, which restricts cliques to those  of sizes $1$ and $2$. Notice that here graph $\mathcal{G}$ can have loops which is different from the graph model in \cite[Algorithm~10]{Mastin:2016}.  In Algorithm~\ref{alg:m}, if $v_t$ is looped, we remove it from the graph and proceed to the next iteration. Otherwise, we try to match it to a previously arrived unmatched vertex. If a match found, we remove both vertices from the graph and proceed to the next iteration. Otherwise, we leave $v_t$ for possible matching in the next iterations. 

The use of Algorithms~\ref{alg:file_plc} and \ref{alg:m} for placement and delivery leads to a method that we call \acrlong{cfcm} (\acrshort{cfcm}).

\section{Performance Analsyis}\label{sec:ana}
In this section, we analyze the expected delivery rates of the proposed file caching schemes through a modeling of the dynamics of Algorithms~\ref{alg:cc} and \ref{alg:m} by differential equations. 

\subsection{Overview of Wormald's Differential Equations Method}\label{subsec:w}
To analyze the performance of \acrshort{cfcm} and \acrshort{cfcc}, we use Wormald's differential equation method and in particular Wormald's theorem \cite[Theorem~5.1]{Wormald:1999}. A formal statement of the theorem is provided in Appendix~\ref{app:w}. Here, we present a rather qualitative description of the theorem and how it can be used to analyze the performance of the greedy procedures on random graphs.

Consider a dynamic random process whose state evolves step by step in discrete time. Suppose that we are interested in the time evolution of some property $P$ of the state of the process. Since the process is random, the state of $P$ is a random variable. To determine how this random variable evolves with time, a general approach is to compute its expected changes per unit time at time $t$, and then regard time $t$ as continuous. This way, one can write the differential equation counterparts of the expected changes per unit of time in order to model the evolution of the variable. Under certain conditions, the random steps follow the expected trends with a high probability, and as a result, the value of the random variable is concentrated around the solution of the differential equations. In this context, Wormald's theorem states that if (i) the change of property $P$ in each step is small, (ii) the rate of changes can be stated in terms of some differentiable function, and (iii) the rate of changes does not vary too quickly with time, then the value of the random variable is sharply concentrated around the (properly scaled) solution of the differential equations at every moment \cite[Section~1.1]{Wormald:1999},\cite[Section~3]{Diaz:2010}.

Notice that both Algorithms~\ref{alg:cc} and \ref{alg:m} can be viewed as dynamic processes on the random side information graph. In Algorithm~\ref{alg:cc},  a vertex arrives at each time and  potentially joins a previously formed clique. In this case, we are interested in the number of cliques at each time instant, as at the end of the process, this equals the number of coded messages that must be sent. In Algorithm~\ref{alg:m}, the arrived vertex might be matched to a previously arrived vertex. Here, the property of interest is the total number of isolated vertices plus matchings made. We apply Wormald's theorem to analyze the behavior of these random quantities and to approximate their expected values at the end of the process.

\subsection{Statistical Properties of the Random Graph Models for Side Information}\label{subsec:asymptotics}
For the analysis of the side information graphs $\mathcal{D}$ and $\mathcal{G}$, it is required to characterize the joint distribution of the edges and the loops for each of these graphs.

In digraph $\mathcal{D}$, every directed edge or loop is  present with the marginal probability of $q$. However, there exist dependencies in the presence of the different edges or loops in $\mathcal{D}$. 

\begin{notation}
We denote by $e_{uv}$ the Bernoulli random variable representing the directed edge from vertex $u$ to vertex $v$. In case of a loop, we have $u=v$.  This random variable takes values 1 and 0 when the edge is present and absent, respectively. Also, we represent by  $\mathcal{E}$ the set of the $K^2$ random variables representing all the edges and the loops of $\mathcal{D}$. Further, by $f_u$ and $\mathcal{F}_{-u}$ we denote the random file requested by vertex $u$ and the random set of distinct files requested by all vertices except vertex $u$, respectively.\footnote{We have $f_u\in \mathcal{F}_{-u}$ if the file requested by vertex $u$ is also requested by other vertices.}
\end{notation}
\begin{remark}\label{rem:dep_struct}
To characterize the dependencies among the edges and the loops of $\mathcal{D}$, consider the random variable $e_{uv}$. The state of the other variables $\mathcal{E}\setminus e_{uv}$ affects the distribution of $e_{uv}$ in the following way:
\begin{itemize}
\item[(i)] The presence (absence) of a directed edge from any vertex to vertex $u$ implies that the file requested by that vertex is (not) available in the cache of $u$. Hence, if $f_v\in\mathcal{F}_{-v}$, the values of the variables in $\mathcal{E}\setminus e_{uv}$ that correspond to the edges that originate from $u$ fully determine $e_{uv}$. Otherwise, the random variables $\mathcal{E}\setminus e_{uv}$ provide information about the storage of the files $\mathcal{F}_{-v}$ in the cache of $u$. Since the cache size is finite, this information affects the probability of the storage of $f_v$ in the cache of $u$ and hence the distribution of $e_{uv}|\mathcal{E}\setminus e_{uv}$ can be different from the marginal distribution of $e_{uv}$.

\item[(ii)] Furthermore, the values of $\mathcal{E}\setminus e_{uv}$ can affect the probability that $f_v\in\mathcal{F}_{-v}$ (see Fig.~\ref{fig:dependency_structure} for an example). Based on point (i), any change in the probability of $f_v\in\mathcal{F}_{-v}$ can directly change the probability of the presence of $f_v$ in the cache of $u$.
\end{itemize}
An example of point (ii) in Remark~\ref{rem:dep_struct} is shown in Fig.~\ref{fig:dependency_structure} for vertices $v$ and $w$. The state of the edges between these two vertices and their loops implies that $f_v\neq f_w$. Consequently,  $e_{uw}=1$ reduces the probability of the presence of $f_v$ in the cache of $u$, or equivalently the probability of $e_{uv}=1$, from the marginal probability $q=\frac{M}{N}$ to $\frac{M-1}{N-1}<q$.
\begin{figure}
\centering
\begin{tikzpicture}[->,>=stealth',shorten >=1pt,auto,node distance=3cm,
                thick,main node/.style={circle,draw}, scale=0.7]

  \node[main node] (2) {$v$};
  \node[main node] (3) [right of=2] {$w$};
  \node[main node] (1) [below right of=2] {$u$};

  \path
    (2) edge [loop above, dashed] node {$e_{vv}=0$}  (2)
        edge [bend right] node {$e_{vw}=1$} (3) 
    (3) edge [loop above, dashed] node {$e_{ww}=0$} (3)
        edge [bend right] node {$e_{wv}=1$} (2)
    (1) edge [bend right] node {$e_{uw}=1$} (3);    
\end{tikzpicture}
\caption{\small Solid (dashed) lines represents the presence (absence) of an edge or loop and no information is available about the presence of the edges and loops that are not shown. This is an illustration of the dependencies among the edges of $\mathcal{D}$. The values of $e_{ij}$, $i,j\in\{v,w\}$ in this example imply that  files $f_v$ and $f_w$ are distinct, as otherwise we must have had $e_{vv}=1$ and $e_{ww}=1$. Further, $e_{uw}=1$ implies that $f_w$ is cached in $u$. As a result, the probability of $e_{uv}=1$, i.e., $f_v$ being in the cache of $u$, becomes $\frac{M-1}{N-1}$. This is different from the marginal probability $\frac{M}{N}$ for $e_{uv}=1$, which shows the dependency among the edges.
 }\label{fig:dependency_structure}
\end{figure}
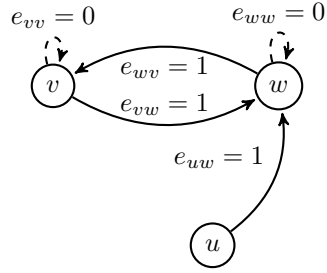
\end{remark}
Because of the dependencies among the edges of $\mathcal{D}$ and therefore the dependencies in the edges of $\mathcal{G}$, an analysis of the performance of \acrshort{cfcm} and \acrshort{cfcc} is difficult. However, we show that in the regime of $\frac{K}{N}\rightarrow 0$ such dependencies become negligible.
\begin{definition}\label{def:asym}
Consider the side information digraph (graph) $\mathcal{X}$ as defined in Definition~\ref{def:digraph} (Definition~\ref{def:graph}) for a network with $K$ caches and a library of $N$ files from which each cache independently and uniformly at random requests a file.
We say a random digraph (graph) $\mathcal{X}_a$ asymptotically approximates $\mathcal{X}$ as $K/N\rightarrow 0$  if it is defined over the same set of vertices as $\mathcal{X}$, and for each pair of vertices $u$ and $v$, we have $\lim_{\frac{K}{N}\rightarrow 0}\mathbb{P}^{(\mathcal{X})}(e_{uv}\mid\mathcal{E}\setminus e_{uv})=\lim_{\frac{K}{N}\rightarrow 0 }\mathbb{P}^{(\mathcal{X}_a)}(e_{uv}\mid\mathcal{E}\setminus e_{uv})$, where the superscripts determine the graph to which  the conditional edge probability corresponds.\footnote{Here again, $e_{uv}$ represents the random variable corresponding to the presence of a directed edge from $u$ to $v$ if $\mathcal{X}$ is a digraph and an edge between $u$ and $v$ if $\mathcal{X}$ is an undirected graph. Also, $\mathcal{E}$ represents the set of random variables $e_{uv}$ defined over all combinations of $u$ and $v$.} For simplicity, we refer to $\mathcal{X}_a$ as an asymptotic model for $\mathcal{X}$.
\end{definition}

\begin{proposition}\label{pr:asymptotic}
For $\frac{K}{N}\rightarrow 0$, the random graph $\mathcal{D}$ is asymptotically approximated by the graph $\mathcal{D}_a$ for which every edge and loop is present independently with probability $q$. Also, graph $\mathcal{G}$ is asymptotically approximated by $\mathcal{G}_a$ for which every edge and loop is present independently with probabilities $q^2$ and $q$, respectively.
\end{proposition}
\begin{IEEEproof}
See Appendix~\ref{app:asymptoticproof}.
\end{IEEEproof}
Based on the chain rule for probabilities and since the limits of the conditional probabilities in Definition~\ref{def:asym} are finite,\footnote{In particular, $\lim_{\frac{K}{N}\rightarrow 0}\mathbb{P}^{(\mathcal{X})}(\mathcal{E})=\lim_{\frac{K}{N}\rightarrow 0}\prod_{i}\mathbb{P}^{(\mathcal{X})}(e_i\mid e_1,\ldots, e_{i-1})=\prod_{i}\lim_{\frac{K}{N}\rightarrow 0}\mathbb{P}^{(\mathcal{X})}(e_i\mid e_1,\ldots, e_{i-1})$, where the edge indices represent an arbitrary ordering of the set of all possible edges and loops of the random graph $\mathcal{X}$.} a random graph $\mathcal{X}$ and an asymptotic approximation to it $\mathcal{X}_a$, have the same joint distribution over their edges as $K/N\rightarrow 0$. As a result, if a property defined over random graph $\mathcal{X}$ is solely a function of its edges, for instance the number of cliques or isolated vertices in $\mathcal{X}$, that property behaves statistically the same when defined over $\mathcal{X}_a$ as $K/N\rightarrow 0$.

\subsection{Analysis of the Expected Rate of \acrshort{cfcm} and \acrshort{cfcc}}\label{subsec:analysis}
Algorithms~\ref{alg:cc} and \ref{alg:m} take a generic side information graph $\mathcal{G}$ as input and output the communication scheme comprising the delivery messages. The number of messages constructed by these algorithms is solely a function of the configuration of the edges of $\mathcal{G}$. Moreover, based on Proposition~\ref{pr:asymptotic}, graph $\mathcal{G}_a$ asymptotically approximates $\mathcal{G}$ in the $K/N\rightarrow 0$ regime. Hence, we perform the analysis of the expected delivery rate of the proposed caching schemes in the $\frac{K}{N}\rightarrow 0$ regime using the asymptotic model $\mathcal{G}_a$ for the side information graph. This approach is mathematically tractable, as unlike in $\mathcal{G}$, all the edges and loops are independent from each other in $\mathcal{G}_a$.

Assuming that the input graph to Algorithms~\ref{alg:cc} and \ref{alg:m} is statistically modeled by random graph $\mathcal{G}_a$, and by applying Wormald's theorem, we get the following two results for the outputs of the two algorithms.
\begin{theorem}\label{th:1}
For Algorithm~\ref{alg:m} with input side information graphs that follow the random graph model $\mathcal{G}_a$, the number of isolated vertices plus the number of matchings made by Algorithm~\ref{alg:m} is
\begin{align*}
\frac{1}{2}\left[K(1-q) -\frac{\log(2-(1-q^2)^{K(1-q)})}{\log(1-q^2)}\!\right]+O(\lambda K)
\end{align*} 
with probability $1-O(\frac{1}{\lambda} e^{-K\lambda^3})$, for any $\lambda>0$.
\end{theorem}
\begin{IEEEproof}
See Appendix~\ref{app:mdproof}.
\end{IEEEproof}
\begin{theorem}\label{th:2}
For Algorithm~\ref{alg:cc} with input side information graphs that follow the random graph model $\mathcal{G}_a$, the number of cliques that cover their vertices using Algorithm~\ref{alg:cc} is
\begin{subequations}\label{eq:sys_cc}
\begin{align}\label{eq:R2}
K \sum_{i=1}^K z_i(1;q) + O(\lambda K),
\end{align} 
with probability $1-O(\frac{K}{\lambda} e^{-K\lambda^3})$, where functions $z_i(x;q),\,i=1,\ldots,K$ are given by the unique solution to the system of differential equations 
\begin{align}\label{eq:diff2}
\Bigg\{
	\begin{array}{ll}
	\frac{dz_i}{dx} \!=\! (1\!-\!q)\!\left[
2g_i(\textbf{z})
\!-\!g_{i+1}(\textbf{z})\!+\! g_{i-1}(\textbf{z})\right];\\
\,z_i(0;q)=0,
	\end{array}
\end{align} 
where 
\begin{align}
g_i(\textbf{z})=\prod_{j=i}^K (1-q^{2j})^{Kz_j(x;q)}
\end{align}
\end{subequations}
and $\textbf{z}=(z_1,\ldots,z_K)$.
\end{theorem}
\begin{IEEEproof}
See Appendix~\ref{app:ccproof}.
\end{IEEEproof}

One can combine Proposition~\ref{pr:asymptotic}, and Theorems~\ref{th:1} and \ref{th:2}, to get approximations to the expected delivery rates of \acrshort{cfcm} and \acrshort{cfcc} as is outlined in the following result.
\begin{corollary}\label{cor:main}
For $0\ll K\ll N$, the expected delivery rates of \acrshort{cfcm} and \acrshort{cfcc} can be approximated by
\begin{align}\label{eq:ER1}
\mathbb{E}\left(R_{m}\right) \approx \frac{1}{2}\left[K(1-q) -\frac{\log(2-(1-q^2)^{K(1-q)})}{\log(1-q^2)}\right]
\end{align} 
and
\begin{align}\label{eq:ER2}
\mathbb{E}\!\left(R_{cc}\right) \approx K \sum_{i=1}^K z_i(1),
\end{align}respectively. 
\end{corollary}
The approximations in (\ref{eq:ER1}) and (\ref{eq:ER2}) become tight as $K\rightarrow\infty$ and $\frac{K}{N}\rightarrow 0$. The former is required for the validity of the concentration results in Theorems~1 and 2.\footnote{This is because the asymptotics denoted by $O$ in Theorems~\ref{th:1} and \ref{th:2} are for $K\rightarrow+\infty$, and the probabilities in the statement of the theorems approach 1 as $K$ increases.} The latter condition ensures that the asymptotic models $\mathcal{D}_a$ and $\mathcal{G}_a$ are valid statistical representations of the side information graphs $\mathcal{D}$ and $\mathcal{G}$ as per Proposition~\ref{pr:asymptotic}. Notice that this condition is satisfied in many practical scenarios where the number of files to be cached is considerably larger than the number of caches in the network. 

\begin{remark}\label{rem:hist}
Applying Algorithm~\ref{alg:cc} to random graph $\mathcal{G}_a$ has the property that as vertex $v_t$ arrives, the behavior of the algorithm up to time $t$ does not provide any information about the connectivity of vertex $v_t$ to the previously arrived vertices. In other words, $v_t$ is connected to any of the previously arrived vertices with probability $q^2$. This is because Algorithm~\ref{alg:cc} operates based on the structure of the edges in the side information graph. If the different edges are present independently,  the history of the process up to time $t$ does not change the probability of the presence of the edges between $v_t$ and the previously arrived vertices $v_1,\ldots v_{t-1}$. This property is essential for the proof of Theorem~\ref{th:2}.
\end{remark}
\begin{remark}\label{rem:nonuni}
In the alternative setting that the popularities of files are nonuniform, the more popular files would be cached with higher probabilities during placement. In that case, the property discussed in Remark~\ref{rem:hist} does not hold anymore for Algorithm~\ref{alg:cc}. In particular, at step $t$, the posterior probability 
of $v_t$ to connect to the previously arrived vertices was strongly dependent on the history of the process up to time $t$.
For instance, the probability of $v_t$ to connect to the vertices that have formed a clique with a large size is higher than the probability of $v_t$ to connect to vertices that belong to small cliques. This is because a vertex that belongs to a large clique is more likely to correspond to a request for a more popular file. Such a file is more likely to be availabe in each cache, including the one that corresponds to vertex $v_t$.
The modeling of the posterior probabilities of the presence of edges makes the analysis of \acrshort{cfcc} difficult for the side information graphs arisen from nonuniform demands.
\end{remark}

\section{Performance Comparison and Simulations}\label{sec:sim}
In this section, we present numerical examples and simulation results for the application of the \acrshort{cfc} schemes that we proposed, to demonstrate the effectiveness of \acrshort{cfc} in reducing the delivery rate. We further show that the  expressions in (\ref{eq:ER1}) and (\ref{eq:ER2}) tightly approximate the expected delivery rates of \acrshort{cfcm} and  \acrshort{cfcc}, respectively.

We use the expected delivery rates of two reference schemes to comment on the performance of our proposed \acrshort{cfc}. The reference cases are the uncoded caching and the optimal decentralized \acrshort{csc} scheme derived in \cite{Maddah_optimal:2016}. The latter is the state-of-the-art result on \acrshort{csc} which provides the optimal memory-rate tradeoff over all coded cachings with uncoded placement.  Although the asymptotic average delivery rate derived in \cite[Theorem~2]{Maddah_optimal:2016} is valid in the infinite file size regime and the underlying caching scheme requires an exponential number of subfiles in $K$,  it provides a suitable theoretical reference for our comparisons. To compute the expected delivery rate of the optimal decentralized \acrshort{csc}, one needs to find the expectation in the RHS of \cite[Theorem~2]{Maddah_optimal:2016}. This is done in Appendix~\ref{app:exp}, where it is shown that
\begin{align}\label{eq:e_r}
\mathbb{E}_\ve{d}(R^*_{\text{CSC}}) =\frac{N-M}{M}
\left[1-\sum_{n=1}^K \frac{\binom{N}{n}\stirling{K}{n}n!}{N^K}(1-M/N)^{n}\right],
\end{align}
where $\stirling{K}{n}$ represents the Stirling number of the second kind \cite[Section~5.3]{Cameron:1994}. 
Also, the expected rate of uncoded caching is derived as
\begin{align}\label{eq:e_r_u}
\mathbb{E}_\ve{d}(R_{\text{uncoded}})&=N\left(1-(1-\frac{1}{N})^K\right)(1-M/N)\\\nonumber
&=K(1-M/N)+O(1/N^2)
\end{align}
in Appendix~\ref{app:exp}.
We use the approximation $\mathbb{E}_\textbf{d}(R_{\text{uncoded}})\approx K(1-M/N)$ throughout this section, as we use $N\geq 100$ in all the following examples.
\begin{figure}
    \centering
    \begin{subfigure}[b]{.45\textwidth}
        \includegraphics[width=.9\textwidth]{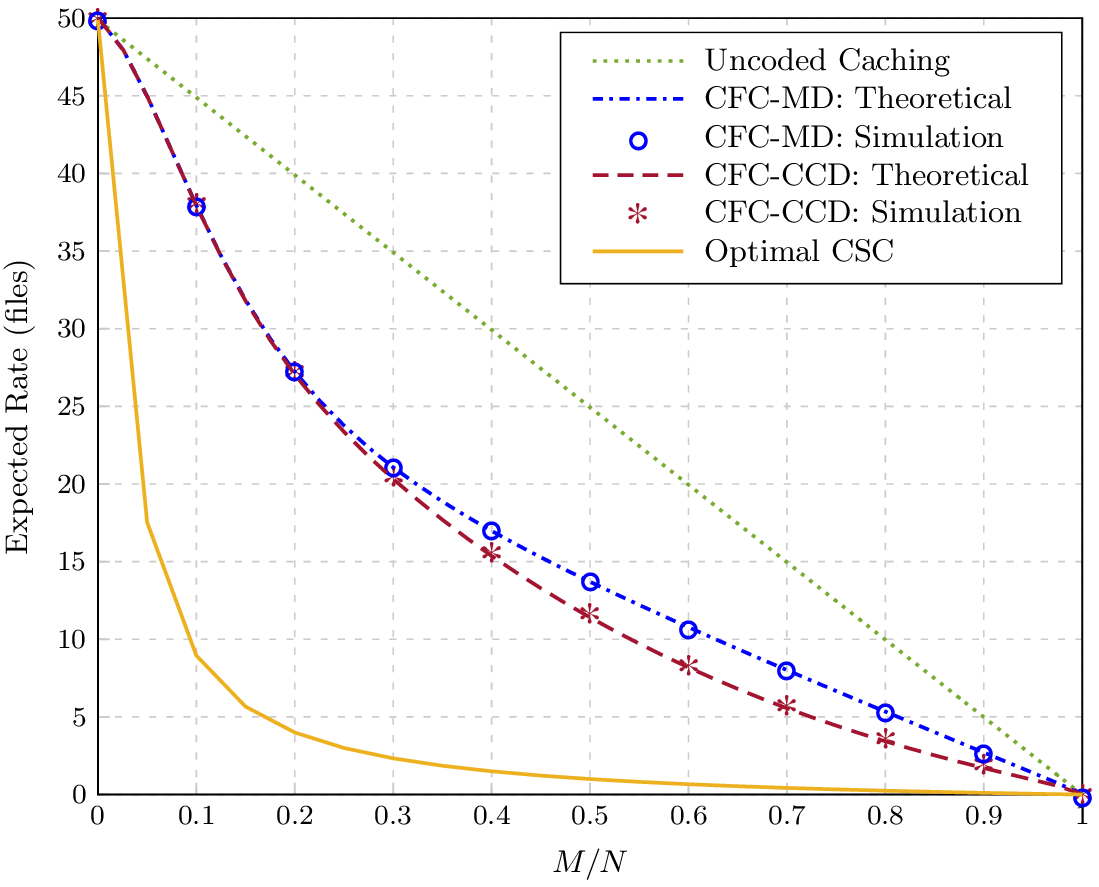}
        \caption{$K=50$, $N=1000$}
        \label{fig:k50}
    \end{subfigure}
    
    \begin{subfigure}[b]{.45\textwidth}
        \includegraphics[width=.9\textwidth]{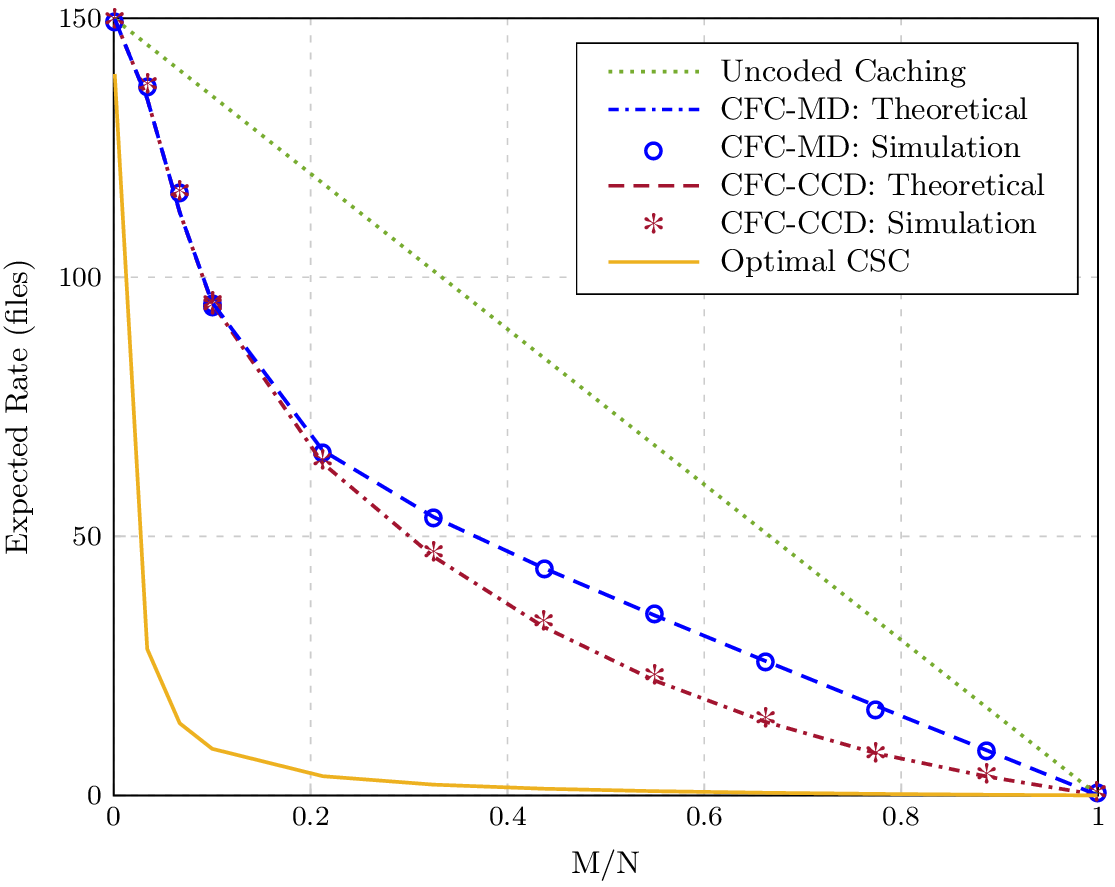}
        \caption{$K=150$, $N=100$}
        \label{fig:k100}
    \end{subfigure}
    \caption{Comparison of the delivery rates of the different caching schemes.}\label{fig:k}
\end{figure}

\subsection{Numerical Examples for Expected Rates}
\begin{figure}
\centering
\begin{minipage}{.45\textwidth}
  \centering
\includegraphics[width=.9\linewidth]{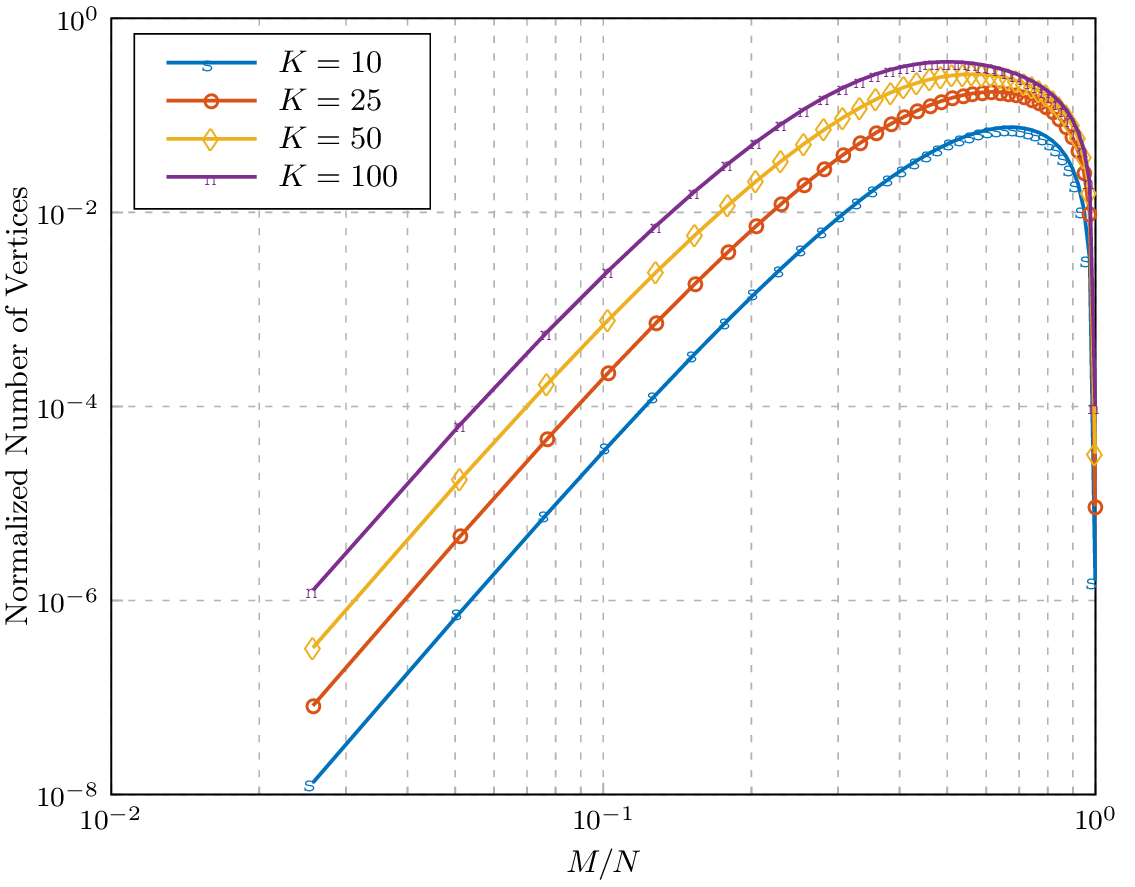}
\caption{Expected number of vertices that are covered by cliques of sizes larger than 2 by Algorithm~\ref{alg:cc}, normalized by $K$.  Here, $N=1000$.}\label{fig:m_justi}
\end{minipage}\qquad
\begin{minipage}{.45\textwidth}
  \centering
  \includegraphics[width=.88\linewidth]{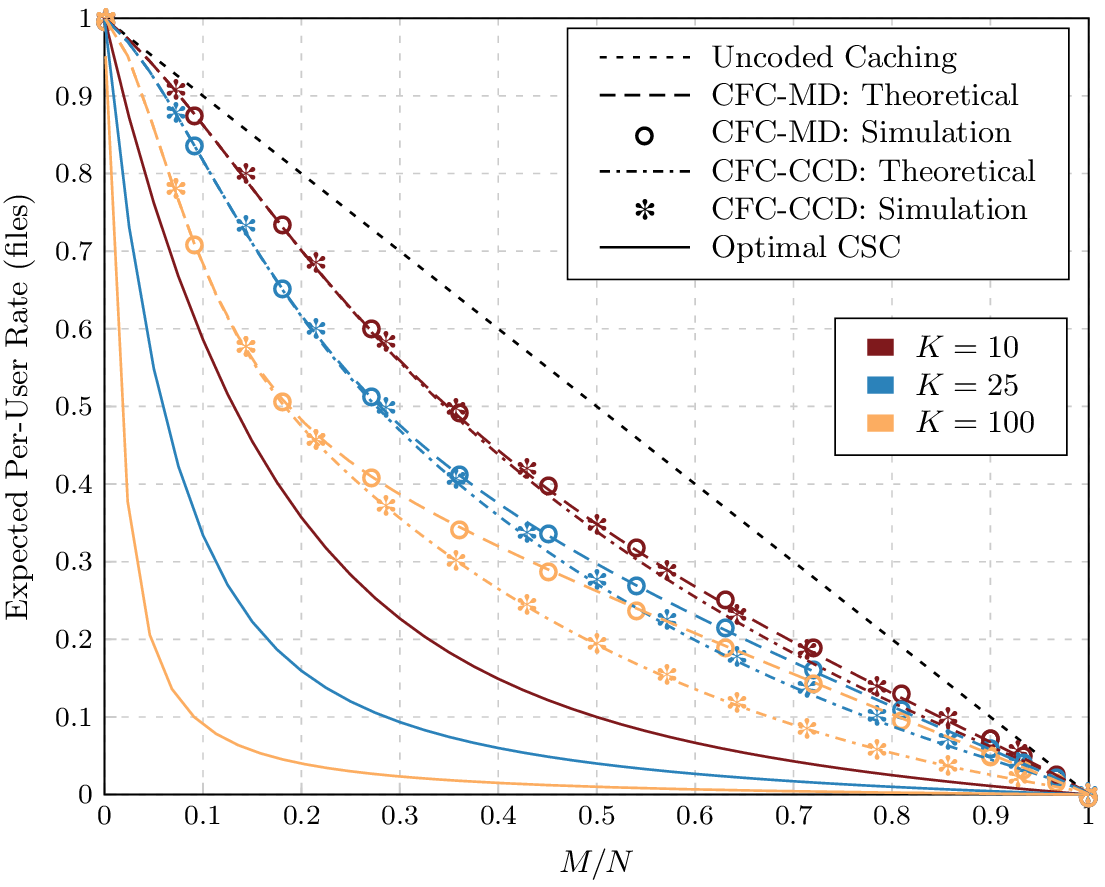}
  \captionof{figure}{The expected per-user delivery rates for caching networks with different numbers of caches. Here, $\frac{K}{N}=0.1$.}
  \label{fig:allK1}
\end{minipage}
\end{figure}

Fig.~\ref{fig:k} shows the expected delivery rates of \acrshort{cfcm} and \acrshort{cfcc}, as well as the average rates of the optimal \acrshort{csc} and the uncoded caching for two networks, one with parameters $K=50$ and $N=1000$ and the other with parameters $K=150$ and $N=100$. 
First, we observe that \acrshort{cfcc} is  notably effective in reducing the delivery rate. In particular, it reduces the delivery rate by 60 to 70 percent compared to uncoded caching.
Notice that as argued in Section~\ref{subsec:del}, for small cache sizes, the expected delivery rate of \acrshort{cfcm} is close to that of \acrshort{cfcc},  as only a small fraction of vertices are covered by cliques of sizes larger than 2. This is shown in Fig.~\ref{fig:m_justi}. As the cache sizes get larger, the probability of formation of larger cliques increases and therefore \acrshort{cfcc} considerably outperforms \acrshort{cfcm}. 
Hence, in the small-cache regime, the use of \acrshort{cfcm} is practically preferred to \acrshort{cfcc} because of its lower computational complexity. 

Further, we observe that the theoretical expressions in (\ref{eq:ER1}) and (\ref{eq:ER2}) are in agreement with the empirical average rates obtained by simulations. Particularly, we have $\frac{K}{N}=0.05$ and $1.5$ in Figs.~\ref{fig:k50} and \ref{fig:k100}, respectively. The results for the latter case imply that the condition $\frac{K}{N}\rightarrow 0$ required in our theoretical analysis can be too conservative in practice.

In Fig.~\ref{fig:allK1}, the per user expected delivery rates, i.e., the expected delivery rates normalized by $K$, are shown for different numbers of caches. In all cases, $N$ is chosen such that $\frac{K}{N}=0.1$. Notice that although the asymptotics in the rate expressions in Theorems~\ref{th:1} and \ref{th:2} are for $K\rightarrow\infty$, the proposed approximations closely match the simulation results for $K$ as small as 10.  In general, our simulation explorations show that in the regime of $K\geq 20$ and $K/N\leq 0.5$, the approximations in Corollary~\ref{cor:main} are tight.

\subsection{Characterization of Coding Gain}\label{subsec:gain}
We now look at the additive and multiplicative coding gains defined as $g_a=\frac{\mathbb{E}(R_{\text{uncoded}})-\mathbb{E}(R_{\text{coded}})}{\mathbb{E}(R_{\text{uncoded}})}$ and $g_m=\frac{\mathbb{E}(R_{\text{uncoded}})}{\mathbb{E}(R_{\text{coded}})}$, respectively, for \acrshort{cfcc} and \acrshort{cfcm}. Notice that $g_a=1-1/g_m$.
\begin{figure}
    \centering
    \begin{subfigure}[b]{0.45\textwidth}
        \includegraphics[width=.8\textwidth]{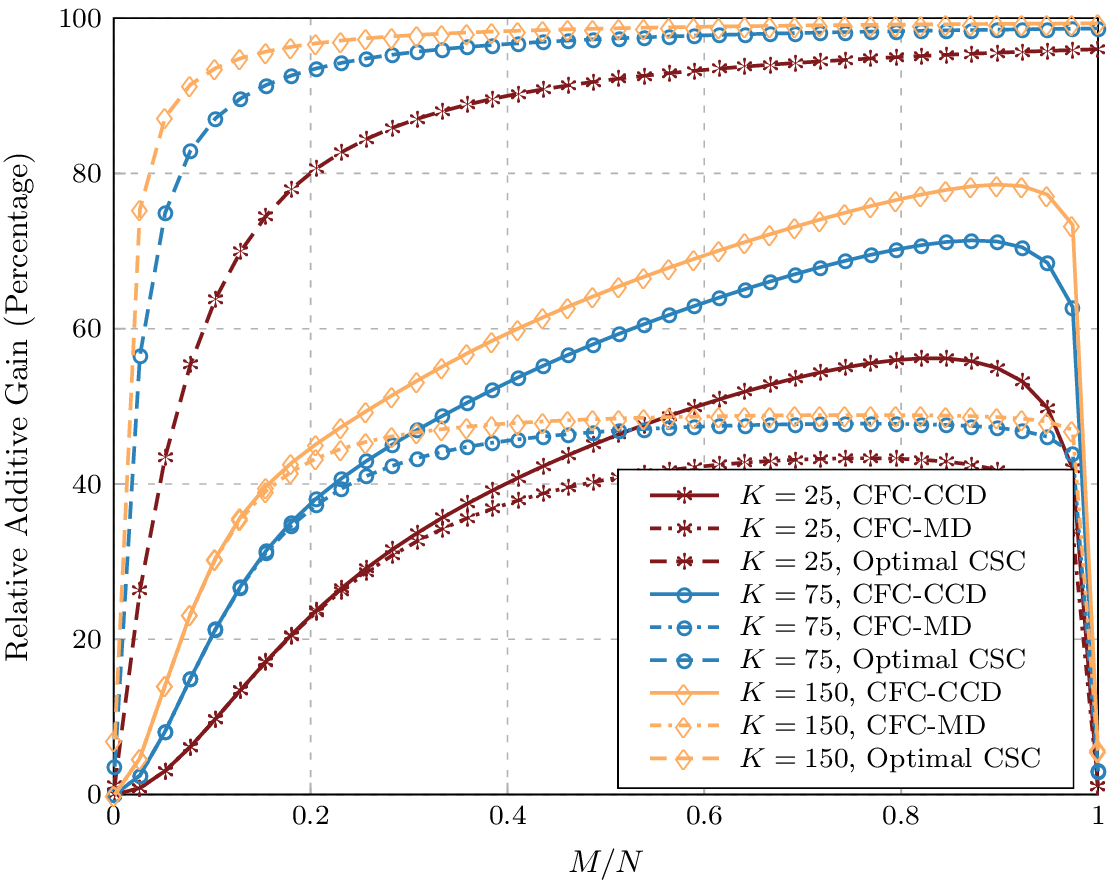}
        \caption{Relative Additive Gain}
        \label{fig:gain_q_add}
    \end{subfigure}
        
    \begin{subfigure}[b]{0.45\textwidth}
        \includegraphics[width=.8\textwidth]{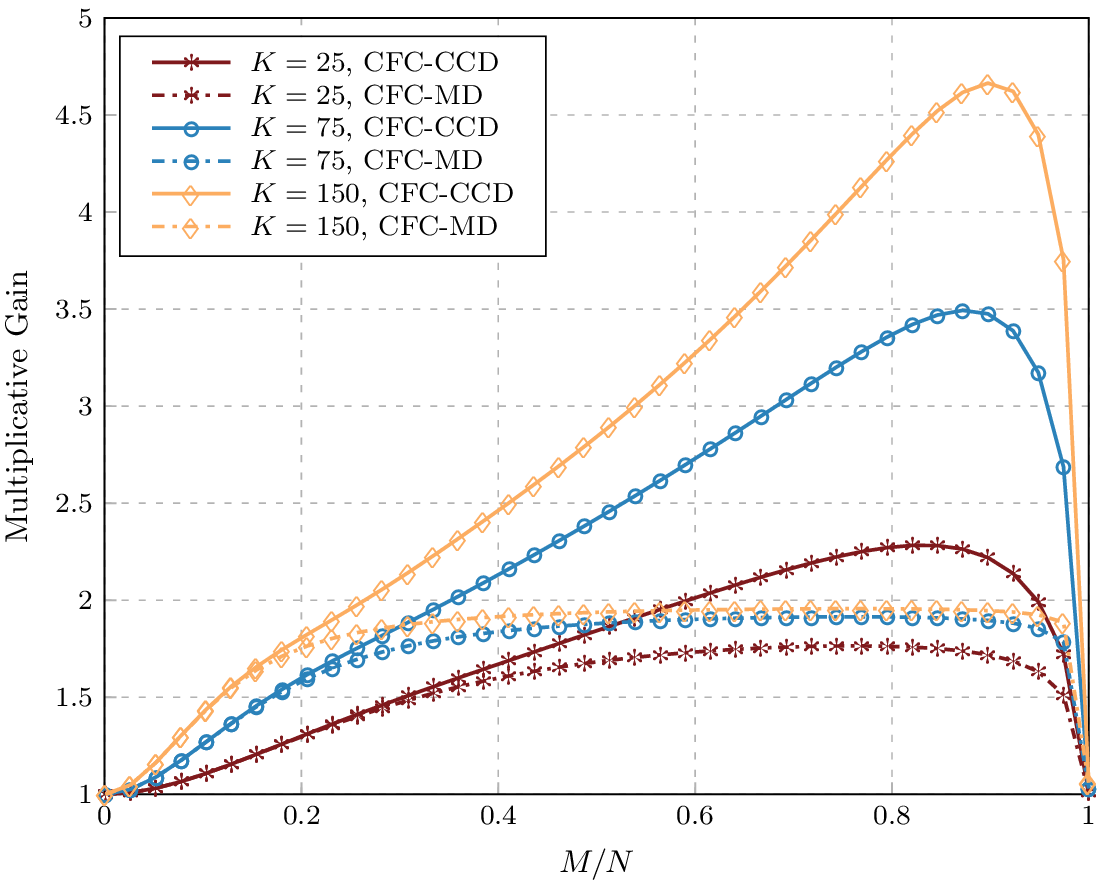}
        \caption{Multiplicative Gain}
        \label{fig:gain_q_mul}
    \end{subfigure}
    
        \begin{subfigure}[b]{0.45\textwidth}
        \includegraphics[width=.8\textwidth]{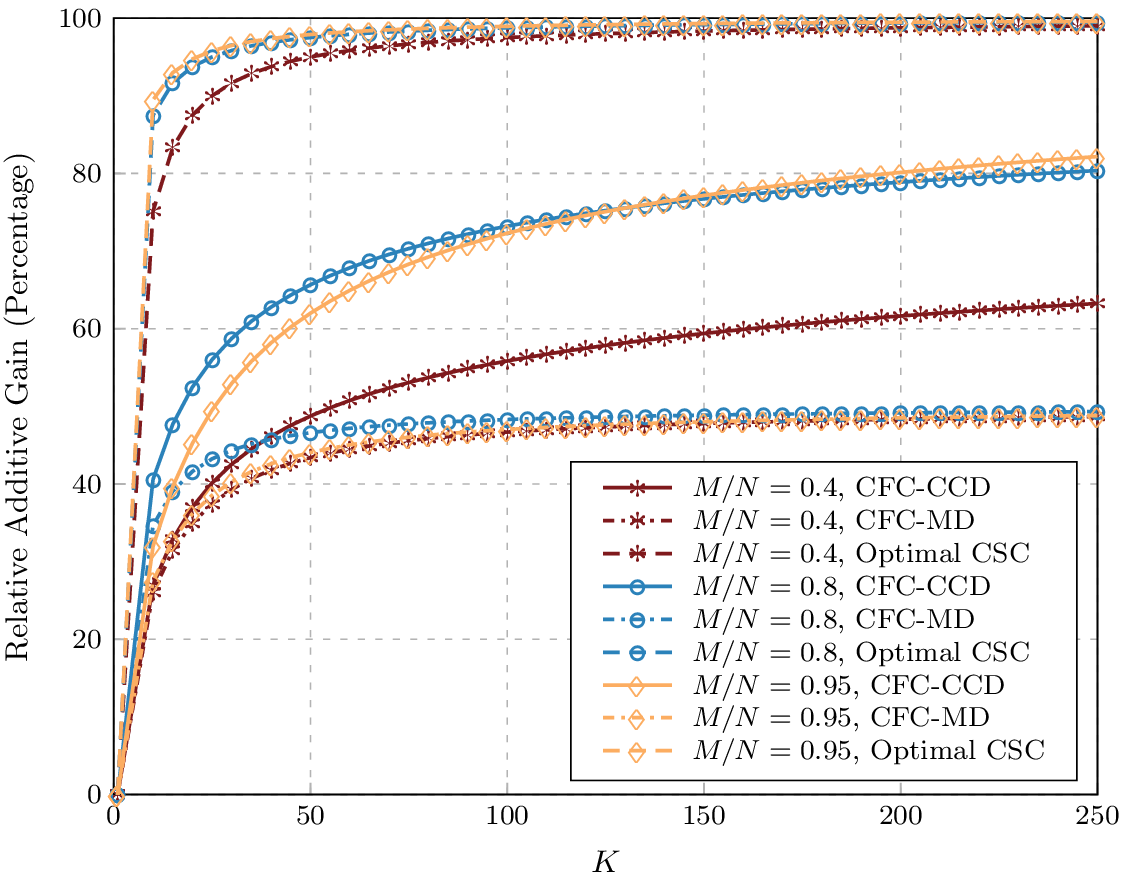}
        \caption{Relative Additive Gain}
        \label{fig:gain_K_add}
    \end{subfigure}
        
    \begin{subfigure}[b]{0.45\textwidth}
        \includegraphics[width=.8\textwidth]{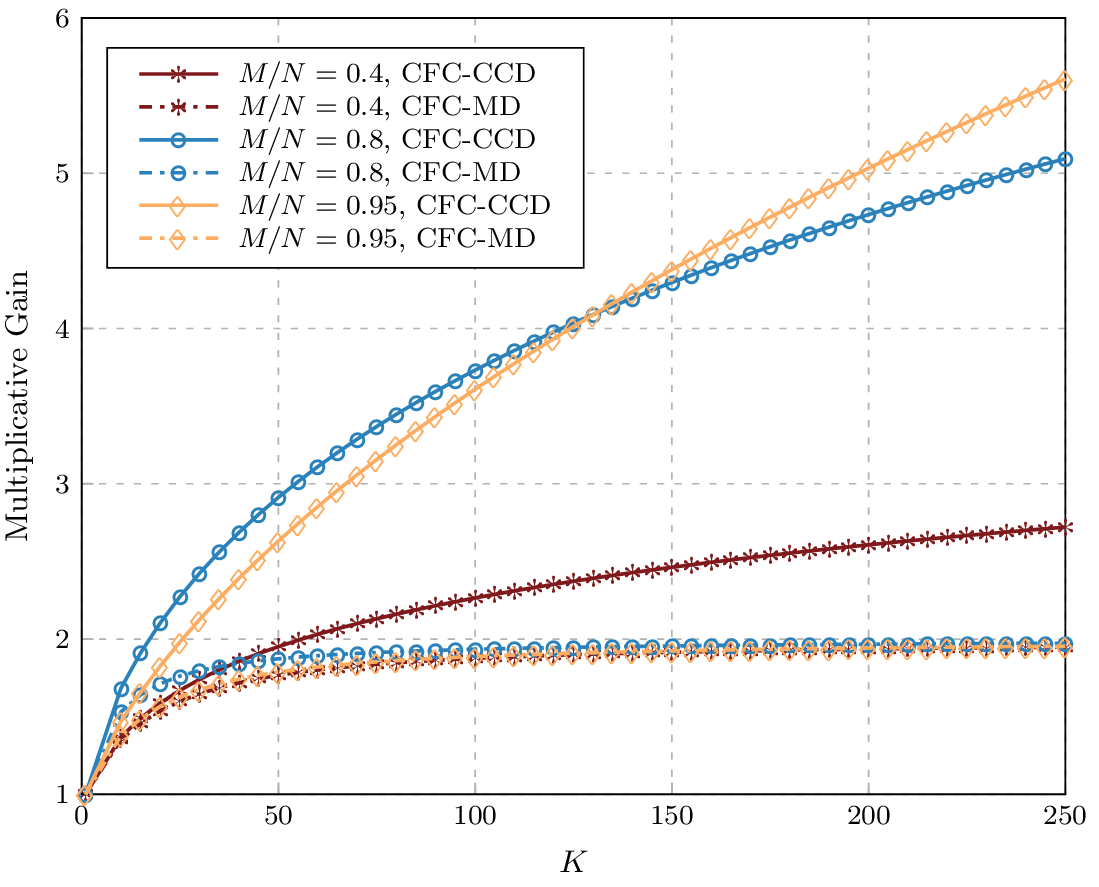}
        \caption{Multiplicative Gain}
        \label{fig:gain_K_mul}
    \end{subfigure}
    \caption{Comparison of the additive and multiplicative coding gains obtained by the different caching schemes. Here,  $N = 100$.}\label{fig:gain}
\end{figure} 
Fig.~\ref{fig:gain} shows $g_a$ and $g_m$ for different coded caching schemes. First, notice that in all cases, \acrshort{cfcc} significantly outperforms uncoded caching. Second, for \acrshort{cfcc}, $g_a$ and $g_m$ reach their maximum for large values of $M/N$, where the gap between the additive gains of \acrshort{cfcc} and optimal subfile caching shrinks significantly. The initial increase of the coding gains with $M/N$ is due to formation of larger cliques in the side information graph.  However, as $M\rightarrow N$, the gains decrease again. This behavior is due to the fact that although for large $M/N$ most of the vertices are connected, at the same time most of the vertices are also looped and are removed by Algorithm~\ref{alg:cc}. Hence, for $M\approx N$, the chance of formation of large cliques diminishes, causing $g_m$ to approach $1$ and consequently $g_a$ to drop to $0$.
Third, the coding gains increase with $K$ because of the higher number of larger cliques formed in the side information graph. For \acrshort{cfcm}, $g_m$ is upper bounded by 2 which is the multiplicative coding gain of \acrshort{cfcm} when  perfect matching occurs. Fourth, notice the gap of the curves for the optimal \acrshort{csc} to 100\% at $M/N=1$. This gap is equal to $\frac{1}{N(1-(1-1/N)^K)}\times 100\%$\footnote{This is a direct consequence of eqs.~(\ref{eq:ave_uncoded}) and (\ref{eq:ave_rate_csc}) in Appendix~\ref{app:exp}  which imply that
$
\frac{\mathbb{E}(R_{\text{uncoded}})}{\mathbb{E}(R^*_{\text{CSC}})}
=\frac{N(1-(1-1/N)^K)q}{\sum_{m=1}^K\mathbb{P}(N_{\text{e}}(\ve{d})=m)(1-(1-q)^m)}.
$
As a result,
$
\lim_{q\rightarrow 1} \frac{\mathbb{E}(R_{\text{uncoded}})}{\mathbb{E}(R^*_{\text{CSC}})}= N(1-(1-1/N)^K)
$ and $\lim_{q\rightarrow 1}\frac{\mathbb{E}(R_{\text{uncoded}})-\mathbb{E}(R^*_{\text{CSC}})}{\mathbb{E}(R_{\text{uncoded}})}=1-\frac{1}{N(1-(1-1/N)^K)}$. }, which can be approximated by $\frac{1}{K}\times 100\%$ for large $N$.

Based on our observation in Fig.~\ref{fig:gain}, the multiplicative gain of \acrshort{cfcc} is limited compared to the optimal \acrshort{csc}, as in the latter, it grows almost linearly with $Kq$.\,\footnote{To keep the other curves readable, the multiplicative coding gain of the optimal \acrshort{csc} is not shown in Figs.~\ref{fig:gain_q_mul} and \ref{fig:gain_K_mul}.}  However, we observe that a considerable portion of the additive gain of optimal \acrshort{csc} can be revived by \acrshort{cfcc}. Hence, it is better justified to look at the coding gains of the proposed \acrshort{cfc} schemes as additive and not multiplicative gains.

\section{Extension to Subfile Caching}\label{sec:dis}
We have shown that the proposed \acrshort{cfc} is an easy to implement yet effective technique to reduce the average delivery rate of caching networks. However,  there is a gap between the delivery rates of  the proposed \acrshort{cfc} and the optimal \acrshort{csc}, as the latter allows for an arbitrarily high complexity  in terms of the number of subfiles used per file and also assumes infinite number of packets per file. It is of practical interest to explore the improvement in the expected delivery rate in a scenario where the system can afford a given level of complexity in terms of the number of subfiles used per file.
In this section, we consider this problem and show that the placement and delivery proposed in Algorithms~\ref{alg:file_plc} and \ref{alg:cc} also provide a framework for \acrshort{csc} when a small number of subfiles per file is used for caching. The resulting \acrshort{csc} scheme provides a means to tradeoff between the delivery rate and the implementation complexity, i.e., the number of subfiles used per file.

Throughout this section, performance evaluations are based on computer simulations as the theoretical results of Section~\ref{sec:ana} cannot be extended to subfile caching. 

\subsubsection{Placement}
For subfile caching, we  break each file in the library  into $\Delta\geq 1$ subfiles.  This leads to a library of $N\Delta$ subfiles, where each subfile is of length $F/\Delta$. Then, we apply the placement in Algorithm~\ref{alg:file_plc} to the library of subfiles. Notice that the proposed subfile placement is different from the decentralized placement of \cite{Maddah_decentralized:2014}. In particular, here all the subfiles are of the same length. Moreover, in the placement of \cite{Maddah_decentralized:2014}, the number of packets of each file stored in each cache was $M/NF$, while this number is random here. Also notice that,  for each cache, the prefetching of the subfiles that belong to different files are dependent, as the total size of the cached subfiles must be at most $M$. 
\begin{remark}
The choice of $\Delta$ depends on the level of complexity that is tolerable to the system. The only restriction that the proposed placement imposes on $\Delta$ is for the ratio $F/\Delta$ to be an integer value. This condition can be easily satisfied in the finite file size regime.
\end{remark}
\subsubsection{Delivery}
Similar to the delivery of file caching, for subfile caching the server forms the side information graphs $\mathcal{D}$ and $\mathcal{G}$ upon receiving the user demands. Then, it delivers the requests using Algorithm~\ref{alg:cc}, by treating each subfile like a file.  Notice that we do not consider \acrshort{csc} with online matching delivery in this section, as its coding gain is limited to 2 and this bound does not improve by increasing the number of subfiles.
\paragraph*{Side information graphs} For subfile caching, both the side information graphs $\mathcal{D}$ and $\mathcal{G}$ have $K\Delta$ vertices. Similar to the approach we followed in Proposition~\ref{pr:asymptotic}, in the following analysis, we consider $\mathcal{D}_a$ and $\mathcal{G}_a$ as the asymptotic counterparts of $\mathcal{D}$ and $\mathcal{G}$ for $\frac{K\Delta}{N\Delta}\rightarrow 0$. Let $w_{k,\delta}$ represent the vertex corresponding to cache $k$ and subfile $\delta$. In $\mathcal{D}_a$, vertex $w_{k,\delta}$ has a loop if subfile $\delta$ of the file requested by cache $k$ is available in cache $k$. The probability of a loop is $q$. The main structural difference between graphs $\mathcal{D}_a$ for file and subfile caching is that the presence of the edges are highly dependent in subfile caching. In particular, for two given caches $k$ and $l\neq k$, and a given subfile $\delta$ of the file requested by cache $k$,  there exist $\Delta$ directed edges from $w_{l,\theta}$  to $w_{k,\delta}$ for all $\theta=1,\ldots,\Delta$, if subfile $\delta$ of the file requested by cache $k$ is available in cache $l$. This event has probability $q$. Otherwise, \textit{all}  $\Delta$ edges from $w_{l,\theta}$  to $w_{k,\delta}$ with $\theta=1,\ldots,\Delta$  are absent.  Also, for subfile caching, we draw no edges from  $w_{k,\delta}$ to $w_{k,\theta}$ for $\delta\neq \theta$. This does not affect the output of the delivery algorithm as if subfile $\theta$ of file $X_k$ is present in cache $k$, vertex $w_{k,\theta}$ is looped and hence is ignored (not joined to any clique) by Algorithm~\ref{alg:cc}.
The latter two properties make the model of the random graph $\mathcal{D}_a$ for subfile caching significantly different from the model used for file caching.

Graph $\mathcal{G}_a$ is built from $\mathcal{D}_a$ in exactly the same way as for file caching. As a result, each vertex is looped with probability $q$. Also, any two vertices belonging to two different caches are present with marginal probability of $q^2$. However, the presence of the edges between the vertices of two caches are highly dependent because of the discussed structures in $\mathcal{D}_a$.
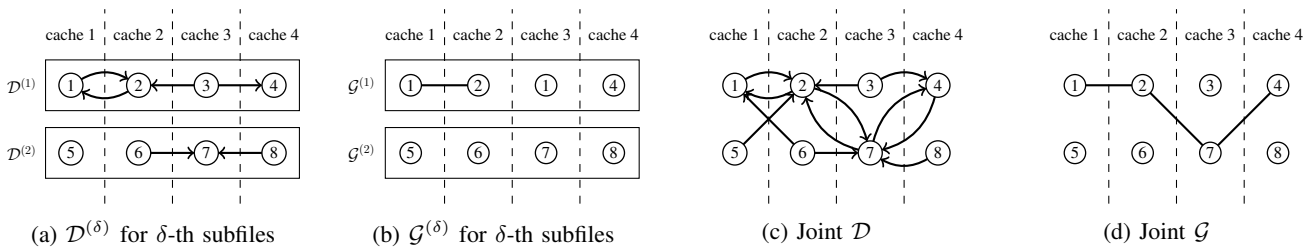
\begin{figure}[t!]
\centering
\begin{subfigure}[t]{0.23\textwidth}
\centering
\tikzset{
  VertexStyle/.append style = { inner sep=3pt},
  EdgeStyle/.append style = {->, bend left} }
\begin{tikzpicture}[scale=.45,transform shape]
{\Large

  \SetGraphUnit{2}  
  \Vertex{3}
  \WE(3){2}
  \WE(2){1}
  \EA(3){4}
  \begin{scope}[VertexStyle/.append style = {minimum size = 5pt, 
   inner sep = 0pt,
   draw=none}]

 \end{scope}
  \Edge(1)(2)
  \Edge(2)(1)
    \tikzset{EdgeStyle/.append style = {bend left = 0}}

  \Edge(3)(4)
  \Edge(3)(2)

  \SO(1){5}
  \EA(5){6}
  \EA(6){7}
  \EA(7){8}
  \begin{scope}[VertexStyle/.append style = {minimum size = 5pt, 
   inner sep = 0pt,
   draw=none}]
 
 \end{scope}
  \Edge(6)(7)
  \Edge(8)(7) 
  \node at (-5.45,0) {$\mathcal{D}^{(1)}$}; 
  \draw[] (-4.75,-0.75) rectangle (2.75,.75);
  \node at (-5.45,-2) {$\mathcal{D}^{(2)}$};
  \draw[] (-4.75,-2.75) rectangle (2.75,-1.25);
   
  \node at (-4,1.5) {cache 1};  
  \draw[dashed] (-3,-3.5) -- (-3,2.25);
  \node at (-2,1.5) {cache 2};  
  \draw[dashed] (-1,-3.5) -- (-1,2.25);
  \node at (-0,1.5) {cache 3};  
  \draw[dashed] (1,-3.5) -- (1,2.25);
  \node at (2,1.5) {cache 4};  
  }
\end{tikzpicture}
\caption{\small $\mathcal{D}^{(\delta)}$ for $\delta$-th subfiles}\label{fig:e2_d}
\end{subfigure}%
\quad
\begin{subfigure}[t]{0.23\textwidth}
\centering
\tikzset{
  EdgeStyle/.append style = {-,bend left,thick} }
\begin{tikzpicture}[scale=.45,transform shape]
{\Large
  \SetGraphUnit{2}  

  \Vertex{1}
  \WE(3){2}
  \WE(2){1}
  \EA(3){4}
  \tikzset{EdgeStyle/.append style = {bend left = 0}}
  \Edge(1)(2)
  \begin{scope}[VertexStyle/.append style = {minimum size = 5pt, 
   inner sep = 0pt,
   draw=none}]
 \end{scope}
  
  \SO(1){5}
  \EA(5){6}
  \EA(6){7}
  \EA(7){8}
  \begin{scope}[VertexStyle/.append style = {minimum size = 5pt, 
   inner sep = 0pt,
   draw=none}]
 \end{scope}
  \node at (-5.45,0) {$\mathcal{G}^{(1)}$}; 
  \draw[] (-4.75,-0.75) rectangle (2.75,.75);
  \node at (-5.45,-2) {$\mathcal{G}^{(2)}$};
  \draw[] (-4.75,-2.75) rectangle (2.75,-1.25);
  \node at (-4,1.5) {cache 1};  
  \draw[dashed] (-3,-3.5) -- (-3,2.25);
  \node at (-2,1.5) {cache 2};  
  \draw[dashed] (-1,-3.5) -- (-1,2.25);
  \node at (-0,1.5) {cache 3};  
  \draw[dashed] (1,-3.5) -- (1,2.25);
  \node at (2,1.5) {cache 4};  }
\end{tikzpicture}
\caption{\small $\mathcal{G}^{(\delta)}$ for $\delta$-th subfiles}\label{fig:e2_g}
\end{subfigure}
\medskip
\begin{subfigure}[t]{0.23\textwidth}
\centering
\tikzset{
  VertexStyle/.append style = { inner sep=3pt},
  EdgeStyle/.append style = {->, bend left} }
\begin{tikzpicture}[scale=.45,transform shape]
{\Large
  \SetGraphUnit{2}  
  \Vertex{3}
  \WE(3){2}
  \WE(2){1}
  \EA(3){4}
  \begin{scope}[VertexStyle/.append style = {minimum size = 5pt, 
   inner sep = 0pt,
   draw=none}]
 \end{scope}
  \Edge(1)(2)
  \Edge(2)(1)
  \Edge(3)(4)
  
  \SO(1){5}
  \EA(5){6}
  \EA(6){7}
  \EA(7){8}
  \begin{scope}[VertexStyle/.append style = {minimum size = 5pt, 
   inner sep = 0pt,
   draw=none}]
 \WE[L=$\:$](5){00}   
 \end{scope}
  
  \Edge(8)(7)
  \Edge(7)(2)
  \Edge(7)(4)
  
  \Edge(4)(7)
  \Edge(2)(7)
  \tikzset{EdgeStyle/.append style = {bend left = 0}}
  \Edge(5)(2)
  \Edge(6)(1)
  \Edge(3)(2)
  \Edge(6)(7)
  \node at (-4,1.5) {cache 1};  
  \draw[dashed] (-3,-3.5) -- (-3,2.25);
  \node at (-2,1.5) {cache 2};  
  \draw[dashed] (-1,-3.5) -- (-1,2.25);
  \node at (-0,1.5) {cache 3};  
  \draw[dashed] (1,-3.5) -- (1,2.25);
  \node at (2,1.5) {cache 4};  }
\end{tikzpicture}
\caption{\small Joint $\mathcal{D}$}\label{fig:e3_d}
\end{subfigure}%
\quad
\begin{subfigure}[t]{0.23\textwidth}
\centering
\tikzset{
  EdgeStyle/.append style = {-,bend left,thick} }
\begin{tikzpicture}[scale=.45,transform shape]
{\Large
  \SetGraphUnit{2}  

  \Vertex{3}
  \WE(3){2}
  \WE(2){1}
  \EA(3){4}
  \begin{scope}[VertexStyle/.append style = {minimum size = 5pt, 
   inner sep = 0pt,
   draw=none}]
 \WE[L=$\:$](1){0}   
 \end{scope}
  
  \SO(1){5}
  \EA(5){6}
  \EA(6){7}
  \EA(7){8}
  \begin{scope}[VertexStyle/.append style = {minimum size = 5pt, 
   inner sep = 0pt,
   draw=none}]
 \end{scope}
   \tikzset{EdgeStyle/.append style = {bend left = 0}}

 \Edge(1)(2)
 \Edge(2)(7)
 \Edge(4)(7)
  \node at (-4,1.5) {cache 1};  
  \draw[dashed] (-3,-3.5) -- (-3,2.25);
  \node at (-2,1.5) {cache 2};  
  \draw[dashed] (-1,-3.5) -- (-1,2.25);
  \node at (-0,1.5) {cache 3};  
  \draw[dashed] (1,-3.5) -- (1,2.25);
  \node at (2,1.5) {cache 4};  }
\end{tikzpicture}
\caption{\small Joint $\mathcal{G}$}\label{fig:e3_g}
\end{subfigure}
\caption{$K=4$ caches and $\Delta=2$ subfiles per file. (a), (b): Side information graphs for separate delivery of $\delta$-th subfiles with $\delta=1,\ldots,\Delta$. (c), (d): Side information graphs for joint delivery of all subfiles.}\label{fig:e2}
\end{figure}

Notice that forming the joint side information graphs for the delivery of all $\Delta$ subfiles increases the coding opportunities compared to the case with $\Delta$ separate side information graphs for the disjoint delivery of the $\delta$-th subfiles of every file. In other words, the number of edges in the joint side information graph $\mathcal{G}_a$ grows superlinearly in $\Delta$ as each of the $K\Delta$ vertices can connect to $(K-1)\Delta$ other vertices. Fig.~\ref{fig:e2} shows an example of separately formed and the corresponding jointly formed side information graphs.
\begin{remark}
Coding opportunities increase by increasing $\Delta$. Hence, in practice, $\Delta$ is determined by the highest level of complexity that is tolerable to the system in terms of the number of subfiles used. 
\end{remark}

\paragraph*{Delivery Algorithms}
The delivery process is complete with the side information graph $\mathcal{G}$ inputted to Algorithm~\ref{alg:cc}. We call the resulting scheme \acrlong{cscc} (\acrshort{cscc}).

\subsubsection{Simulation Results}
\begin{figure}
\centering
\begin{subfigure}{0.45\textwidth}
  \centering
  \includegraphics[width=.8\linewidth]{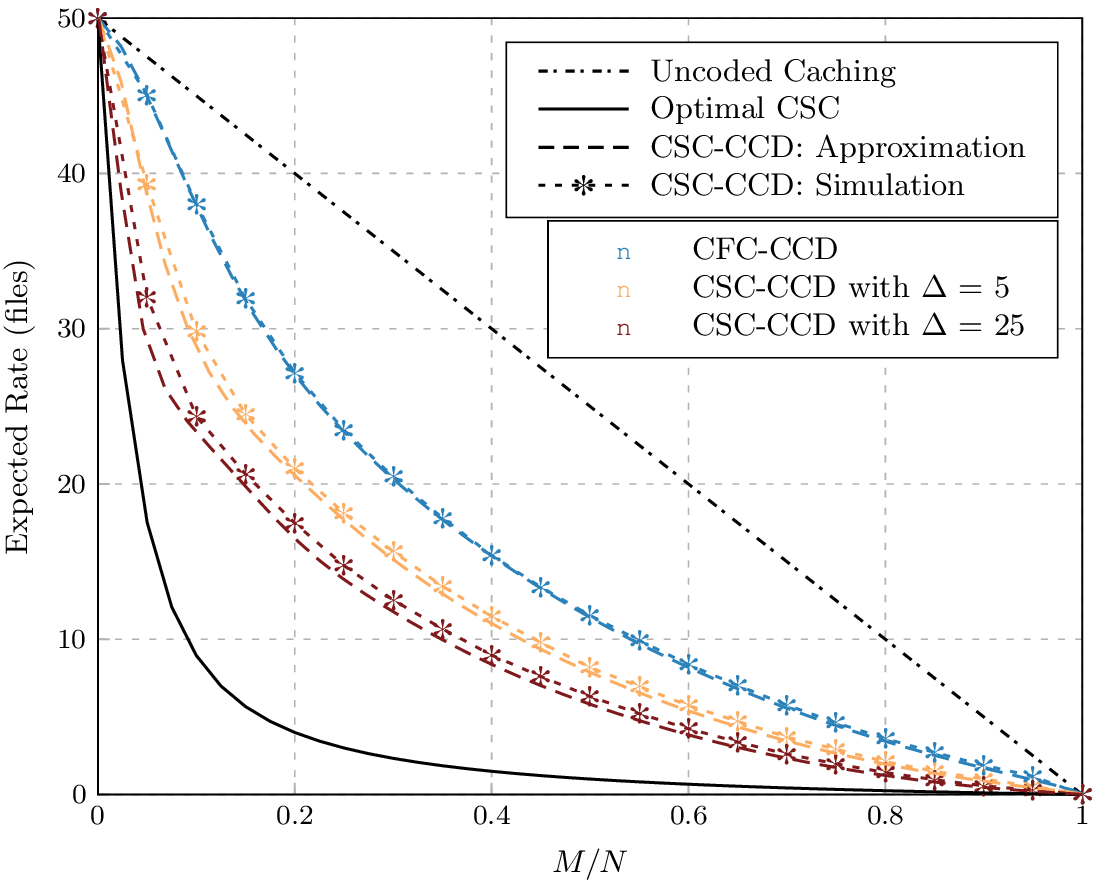}
  \caption{Average Rates}
  \label{fig:sub}
\end{subfigure}

\begin{subfigure}{0.45\textwidth}
  \centering
  \includegraphics[width=.8\linewidth]{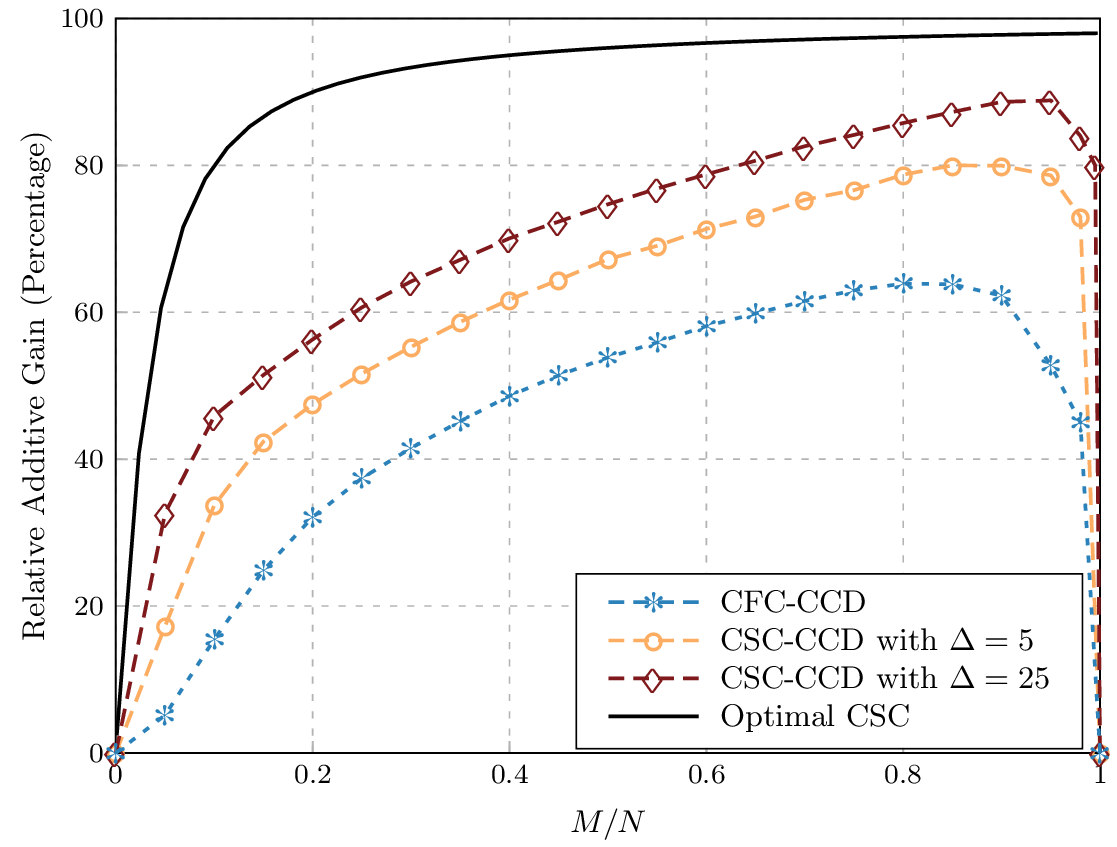}
  \caption{Additive Gains}
  \label{fig:subR}
\end{subfigure}

\begin{subfigure}{0.45\textwidth}
  \centering
  \includegraphics[width=.8\linewidth]{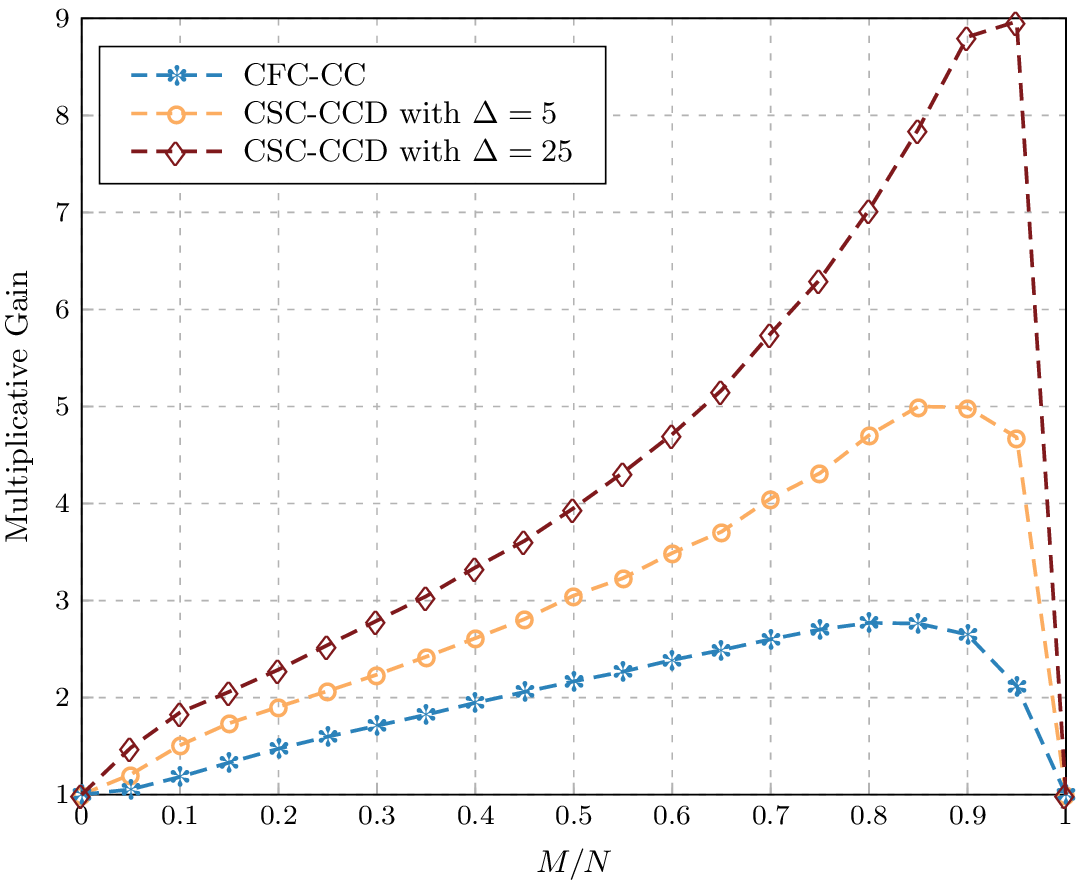}
  \caption{Multiplicative Gains}
  \label{fig:subM}
\end{subfigure}
\caption{\small Figure (a) shows both theoretical approximations and simulation results. Figures (b) and (c) are based on computer simulations. Here $K=50$ caches and $N=1000$ files.}
\end{figure}

For a system with $K=50$ caches, Fig.~\ref{fig:sub} shows the expected delivery rates of \acrshort{cscc} with $\Delta=5$ and $\Delta=25$ as well as the expected delivery rates of \acrshort{cfcc}, the optimal decentralized \acrshort{csc} of \cite{Maddah_optimal:2016} and uncoded caching. Notice that \acrshort{cfcc} is identical to \acrshort{cscc} with $\Delta=1$ subfile. 

An important observation is that relative to the $2^K$ subfiles required by the optimal decentralized \acrshort{csc}, the use of a small number of subfiles in \acrshort{cscc} can significantly shrink the rate gap between the delivery rates of \acrshort{cfcc} and the optimal decentralized \acrshort{csc}. In Fig.~\ref{fig:sub}, we have also shown approximate expected delivery rates based on the theoretical results in Theorem~\ref{th:2}. More specifically, we use the approximation $R_{cc,\text{subfile}}\approx \frac{1}{\Delta}R_{cc}(K\Delta)$. As we discussed earlier, because of the structural differences between the side information graphs for file and subfile caching, such an approximation is generally prone to large errors. However, for $\Delta\ll K$ and $N\Delta\gg K\Delta$, it still results in an approximation for the expected delivery rate.

Figs.~\ref{fig:subR} and \ref{fig:subM} show the additive and multiplicative coding gains for the proposed \acrshort{csc} scheme.  One observes that the rate of change in the additive gain decays as the number of subfiles increases. For instance, increasing the number of subfiles from $\Delta=1$ to $\Delta=5$ results in a larger reduction in the delivery rate compared to the case where the number of subfiles is increased from $\Delta=5$ to $\Delta=25$. However, the multiplicative gain increases significantly when a larger number of subfiles is used.
\section{Conclusion}\label{sec:con}
We explored the decentralized coded caching problem for the scenario where files are not broken into smaller parts (subfiles) during placement and delivery. This scenario is of practical importance because of its simpler implementation. 
We showed that although the requirement of caching the entirety of a file puts restrictions on creation of coding opportunities, coded file caching is still an effective way to reduce the delivery traffic of the network compared to the conventional uncoded caching. In particular, we proposed the \acrshort{cfcc} file caching scheme, which performs delivery based on a greedy clique cover algorithm operating over a certain side information graph.  We further proposed a file caching scheme called \acrshort{cfcm}, which uses a simpler delivery algorithm, and is designed for the small cache size regime. We derived approximate expressions for the expected delivery rate of both schemes.  

Because of the more restricted setup of the file caching problem, the schemes we proposed here are suboptimal compared to the coded subfile cachings, but they still promise considerable gains over uncoded caching. Further, the delivery rate gap between \acrshort{cfcc} and the state-of-the-art decentralized coded subfile cachings shrinks considerably in the regime of large total storage capacity. These findings suggest that in scenarios where the implementation of subfile caching is difficult, it is still possible to effectively gain from network coding to improve the system performance.

Finally, we discussed the generalization of the proposed file caching schemes to subfile caching with an arbitrary number of subfiles per file. We observed that a considerable portion of the performance loss due to the restrictions of file caching can be recovered by using a relatively small number of subfiles per file. While the construction of most of the coded subfile caching methods requires large numbers of subfiles per file, this observation gives grounds for the development of new decentralized subfile cachings that require a small number of subfiles without causing a considerable loss in the delivery rate.

\appendices
\section{Rationale for Choosing the Largest Clique in Algorithm~\ref{alg:cc}}\label{app:lar}
Algorithm~\ref{alg:cc} joins $v_t$ to the \textit{largest} suitable clique at time $t$. The rationale for this choice is as follows. Let $Y_i(t)$ represent the number of cliques of size $i$ right before time $t$, where $i\in\{1,\cdots,K\}$. Let $\mathcal{N}_t$ be the event that there exists no suitable clique for $v_t$ at time $t$. Then,
\begin{align*}
\mathbb{P}(\mathcal{N}_t)=\prod_{i=1}^{K}(1-q^{2i})^{Y_i(t)}.
\end{align*}
Given $Y_i(t)$, and with the knowledge that $v_t$ has joined a clique of size $s$ at time $t$, we get $Y_i(t+1)=Y_i(t),i\not\in\{s,s+1\}$, $Y_s(t+1)=Y_s(t)-1$ and $Y_{s+1}(t+1)=Y_{s+1}(t)+1$. Then, it is easy to show that
\begin{align}\nonumber
\mathbb{P}(\mathcal{N}_{t+1}|v_t\text{ joined a clique of size } s \text{ at time } t,\!\ve{Y}(t))&\\\label{eq:no_clique}=\frac{1-q^{2(s+1)}}{1-q^{2s}}\prod_{i=1}^{K}(1-q^{2i})^{Y_i(t)}&,
\end{align}
where $\ve{Y}(t)=(Y_1(t),\cdots,Y_K(t))$.
In order to minimize the expected rate, one has to make (\ref{eq:no_clique}) small. Now, assume that at time $t$,  multiple suitable cliques existed for $v_t$. It is straightforward to check that (\ref{eq:no_clique}) would take its smallest value if the suitable clique with the largest size $s$ was chosen at time $t$. 
\section{Wormald's Theorem}\label{app:w}
In this appendix, we present Wormald's theorem \cite[Theorem~5.1]{Wormald:1999} and introduce the notations that we use in the proofs of Theorems~\ref{th:1} and \ref{th:2}.\footnote{Wormald's theorem was introduced in \cite[Theorem~1]{Wormald:1995}. The most general setting of the theorem was established in \cite[Theorem~5.1]{Wormald:1999}. A simplified version of the theorem  is provided in \cite[Section~3]{Diaz:2010}, and examples of its application can be found in \cite{Wormald:1999}, \cite[Section~4]{Diaz:2010}, \cite{Yang:2014} and \cite[Section~C.4]{Richardson:2008}.}

Consider a sequence of random processes indexed by $n,\,n=1,2,\ldots\,$.\footnote{For instance, the different processes can be the outcomes of a procedure implemented on graphs with different numbers of vertices $n$. Then, each process associates with one of the graphs.} For each $n$, the corresponding process is a probability space denoted by $(Q_0^{(n)},Q_1^{(n)},\ldots)$, where each $Q_i^{(n)}$ takes values from a set $S^{(n)}$. The elements of the space are $(q_0^{(n)},q_1^{(n)},\ldots)$, where $q_i^{(n)}\in S^{(n)}$. Let $H_t^{(n)}$ represent the random history of process $n$ up to time $t$ and  $h_t^{(n)}$ represent a realization of $H_t^{(n)}$.  Also, we denote by $S^{(n)+}$ the set of all  $h_t^{(n)}=(q_0^{(n)},\ldots,q_t^{(n)})$, where $q_i^{(n)}\in S^{(n)},\,t=0,1,\ldots$. For simplicity of notation, the dependence on $n$  is usually dropped in the following.

Now, consider a set of variables $W_1(t),\ldots,W_a(t)$ defined on the components of the processes. Let $w_i(h_t)$ denote the value of $W_i(t)$ for the history $h_t^{(n)}$. We are interested in analyzing the behavior of these random variables throughout the process.

\begin{theorem}[Wormald's Theorem \textsc{$\!$\cite[Theorem~5.1]{Wormald:1999}}]\label{th:w}
For $1 \leq l \leq a$, where $a$ is fixed, let $w_l: S^{(n)+} \rightarrow \mathbb{R}$ and $f_l: \mathbb{R}^{a+1} \rightarrow \mathbb{R}$, such
that for some constant $c_0$ and all $l$, $|w_l(h_t)| < c_0n$ for all $h_t\in S^{(n)+}$ for all $n$.  Assume the following three conditions hold, where in (ii) and (iii), $D$ is some bounded connected open set containing the closure of
\begin{align*}
\{(0,z_1,\ldots,z_a) : \mathbb{P}(W_i(0) = z_ln, l =1,\ldots, a) = 0\} 
\end{align*}
for some $n$, and $T_D(W_1,\ldots,W_a)$ is the minimum $t$ such that $(t/n, W_1(t)/n, \ldots , W_a(t)/n)\not\in D$.
\begin{itemize}
\item[\textit{(i)}]  (\textit{Boundedness hypothesis}) 
For some functions $\beta = \beta(n) \geq 1$ and $\gamma = \gamma(n)$,  the probability that
\begin{align*}
\max_{i=1,\ldots, a}|W_i(t+1)-W_i(t)|\leq \beta
\end{align*}
conditional upon $H_t$, is at least $1-\gamma$ for $t<T_D$.

\item[\textit{(ii)}] (\textit{Trend hypothesis}) For some function $\lambda_1 = \lambda_1(n) = o(1)$, for all $l \leq a$
\begin{small}
\begin{align*} 
\Big|\mathbb{E}\left(W_l(t\!+\!1)\!-\!W_l(t)|H_t\right)
\!-\!f_l\!\left(\frac{t}{n},\frac{W_1(t)}{n},\ldots, \frac{W_a(t)}{n}\right)\!\Big| \!\leq\! \lambda_1
\end{align*}\end{small}
for $t<T_D$.
\item[\textit{(iii)}]  (\textit{Lipschitz hypothesis}) Each function $f_l$ is continuous and satisfies a Lipschitz condition on $D \cap \{(t, z_1, \ldots , z_a) : t \geq 0\}$, with the same Lipschitz constant for each $l$.
\end{itemize} 

Then, the following are true.
\begin{itemize}
\item[\textit{(a)}] For $(0, \hat{z}_1,\ldots ,\hat{z}_a)\in D$, the system of differential equations defined by
\begin{align*}
\frac{dz_l}{dx} = f_l(x,z_1,\ldots,z_a),\quad l=1,\ldots,a 
\end{align*}
has a unique solution that passes through $z_l(0)=\hat{z}_l,\;l=1,\ldots,a$, which extends to points arbitrarily close to the boundary of $D$.
\item[\textit{(b)}] Let $\lambda > \lambda_1 + c_0 n\gamma$ with $\lambda = o(1)$. For a sufficiently large constant $C$, with probability
$1-O(n\gamma+\frac{\beta}{\lambda} e^{-n\lambda^3/\beta^3})$
\begin{align*}
W_l(t) = nz_l(t/n) + O(\lambda n)
\end{align*}
uniformly for all  $0 \leq t \leq \sigma n$ and for each $l$, where $z_l(x)$ is the solution in (a) with $\hat{z}_l =\frac{1}{n}W_l(0)$, and $\sigma =\sigma(n)$ is the supremum of those x to which the solution can be extended before reaching within $l^\infty$-distance $C\lambda$ of the boundary of $D$.
\end{itemize}
\end{theorem}
Hence, functions $z_l(x)$ model the behavior of $\frac{W_l(nx)}{n}$ for each $n$, and the solution to the system of differential equations provides a deterministic approximation to the dynamics of the process. 
\begin{remark}\label{rem:asy}
In the statement of the theorem, variables $W_l$ and time $t$ are normalized by $n$. This is because in many applications, this normalization leads to only one set of differential equations for all $n$, instead of different systems for each $n$. Also, the asymptotics denoted by $O$ are for $n \rightarrow +\infty$, and the term \textit{``uniformly''} in (b) refers to the convergence implicit in the $O$ terms.
\end{remark}

\begin{remark}\label{rem:W}
A version of Theorem~\ref{th:w} also holds when $a$ is a function of $n$, with the probability in (b) replaced by $1-O(an\gamma+\frac{a\beta}{\lambda} e^{-n\lambda^3/\beta^3})$, under the condition that all functions $f_l$ are uniformly bounded by some Lipschitz constant and $f_l$ depends only on the variables $x,z_1,\ldots,z_l$. The latter condition is because as $n\rightarrow\infty$, one needs to solve a system of infinite number of differential equations which  involves complicated technical issues. However, when $f_l$ depends only on  $x,z_1,\ldots,z_l$, one can solve the finite systems obtained for each $f_l$ by restricting the equations to the ones that involve $x,z_1,\ldots,z_l$.
\end{remark}
 
\section{Proof of Proposition~\ref{pr:asymptotic}}\label{app:asymptoticproof}
We need to prove that for each pair of vertices $u$ and $v$, $\lim_{\frac{K}{N}\rightarrow 0}\mathbb{P}^{(\mathcal{D})}(e_{uv}\mid\mathcal{E}\setminus e_{uv})=\lim_{\frac{K}{N}\rightarrow 0 }\mathbb{P}^{(\mathcal{D}_a)}(e_{uv}\mid\mathcal{E}\setminus e_{uv})$. We have $\lim_{\frac{K}{N}\rightarrow 0 }\mathbb{P}^{(\mathcal{D}_a)}(e_{uv}\mid\mathcal{E}\setminus e_{uv})=\mathbb{P}(e_{uv})=q$ by definition of $\mathcal{D}_a$. Hence, it is enough to prove that $\lim_{\frac{K}{N}\rightarrow 0 }\mathbb{P}^{(\mathcal{D})}(e_{uv}\mid\mathcal{E}\setminus e_{uv})=\mathbb{P}^{(\mathcal{D})}(e_{uv})=q$  for graph $\mathcal{D}$, which implies that $e_{uv}$ is independent of $\mathcal{E}\setminus e_{uv}$ in the $\frac{K}{N}\rightarrow 0$ regime. In the sequel, all probabilities correspond to random graph $\mathcal{D}$. Hence, we write $\mathbb{P}$ in place of $\mathbb{P}^{(\mathcal{D})}$ to minimize notational clutter.

As per point (i) of Remark~\ref{rem:dep_struct}, the  value of $e_{uv}$ is fully determined by the state of the other edges that originate from $u$, if $f_v\in\mathcal{F}_{-v}$.  Using the Baye's rule and the law of total probability for conditioning on $f_v\in\mathcal{F}_{-v}$, we have:
\begin{align}
\mathbb{P}(e_{uv}|\mathcal{E}\setminus e_{uv})&=\frac{\mathbb{P}(e_{uv},\mathcal{E}\setminus e_{uv})}{\mathbb{P}(\mathcal{E}\setminus e_{uv})
}\label{eq:lim_cond0}
\end{align}
where:
\begin{align*}
\mathbb{P}(e_{uv},\mathcal{E}\setminus e_{uv})&=\mathbb{P}(e_{uv},\mathcal{E}\setminus e_{uv}|f_v\in\mathcal{F}_{-v})
\mathbb{P}(f_v\in\mathcal{F}_{-v})\\
&\quad+\mathbb{P}(e_{uv},\mathcal{E}\setminus e_{uv}| f_v\not\in\mathcal{F}_{-v})
\mathbb{P}(f_v\not\in\mathcal{F}_{-v})
\end{align*}
and
\begin{align*}
\mathbb{P}(\mathcal{E}\setminus e_{uv})&= \mathbb{P}(\mathcal{E}\setminus e_{uv}| f_v\in\mathcal{F}_{-v})
\mathbb{P}(f_v\in\mathcal{F}_{-v})\\
&\quad+\mathbb{P}(\mathcal{E}\setminus e_{uv}| f_v\not\in\mathcal{F}_{-v})
\mathbb{P}(f_v\not\in\mathcal{F}_{-v}).
\end{align*}
However, $|\mathcal{F}_{-v}|\leq K-1$. As a result, $\mathbb{P}(f_v\in\mathcal{F}_{-v})\leq \frac{K-1}{N}$ and  $\lim_{\frac{K}{N}\rightarrow 0}\mathbb{P}(f_v\in\mathcal{F}_{-v})=1-\lim_{\frac{K}{N}\rightarrow 0}\mathbb{P}(f_v\not\in\mathcal{F}_{-v})=0$. Hence:
\begin{align}\nonumber
\lim_{\frac{K}{N}\rightarrow 0}\mathbb{P}(e_{uv}|\mathcal{E}\setminus e_{uv})&=\frac{\lim_{\frac{K}{N}\rightarrow 0}
\mathbb{P}(e_{uv},\mathcal{E}\setminus e_{uv}| f_v\not\in\mathcal{F}_{-v})}{\lim_{\frac{K}{N}\rightarrow 0}
\mathbb{P}(\mathcal{E}\setminus e_{uv}| f_v\not\in\mathcal{F}_{-v})}\\\label{eq:lim_cond}
&=\lim_{\frac{K}{N}\rightarrow 0}
\mathbb{P}(e_{uv}|\mathcal{E}\setminus e_{uv}, f_v\not\in\mathcal{F}_{-v}).
\end{align}

By construction, we have
$
\mathbb{P}(e_{uv}\mid\mathcal{E}\setminus e_{uv}, f_v\not\in\mathcal{F}_{-v})=
\mathbb{P}(f_v\in\mathcal{C}_u|\mathcal{E}\setminus e_{uv}, f_v\not\in\mathcal{F}_{-v})
$, 
where $\mathcal{C}_u$ is the set of files stored in the cache of vertex $u$. Since $f_v\not\in\mathcal{F}_{-v}$, based on point (i) in Remark~\ref{rem:dep_struct}, $\mathcal{E}\setminus e_{uv}$  affects the probability of  $f_v\in C_u$ by providing information about the amount of storage used for the files requested by the other caches. This effect is negligible when $\frac{K}{N}\rightarrow 0$. To prove this, we derive a lower and an upper bound on $\mathbb{P}(f_v\in\mathcal{C}_u\mid\mathcal{E}\setminus e_{uv}, f_v\in\mathcal{F}_{-v})$ by considering the two extreme cases where all the other edges and loops that originate from $u$ are present and absent, respectively. Assume that the number of distinct files requested by the vertices other than $v$ is $\alpha$. This implies $\alpha\leq K-1$. Then, the two extreme cases result in the inequalities $\frac{M-\alpha}{N-\alpha}\leq\mathbb{P}(f_v\in\mathcal{C}_u\mid\mathcal{E}\setminus e_{uv}, f_v\in\mathcal{F}_{-v})\leq\frac{M}{N-\alpha}$. Maximizing and minimizing the upper and the lower bounds respectively over alpha, we bound the range of $\mathbb{P}(f_v\in\mathcal{C}_u\mid\mathcal{E}\setminus e_{uv}, f_v\in\mathcal{F}_{-v})$ over all possible demands as 
$\frac{M-(K-1)}{N-(K-1)}\leq\mathbb{P}(f_v\in\mathcal{C}_u\mid\mathcal{E}\setminus e_{uv}, f_v\in\mathcal{F}_{-v})\leq\frac{M}{N-1}$.
Taking the limits of the lower and the upper bounds as $\frac{K}{N}\rightarrow 0$ and using $M=qN$ for a fixed $q$, we get:
\begin{align}\nonumber
\lim_{\frac{K}{N}\rightarrow 0}\mathbb{P}(e_{uv}\mid\mathcal{E}\setminus e_{uv}, f_v\not\in\mathcal{F}_{-v})&=\frac{M}{N}=q=\mathbb{P}(e_{uv})\\\label{eq:lim_cond2}
&=\lim_{\frac{K}{N}\rightarrow 0}\mathbb{P}(e_{uv}\mid f_v\not\in\mathcal{F}_{-v}),
\end{align}
where for the last equality, we used $q=\mathbb{P}(e_{uv})=\lim_{\frac{K}{N}\rightarrow 0}\mathbb{P}(e_{uv}\mid f_v\not\in\mathcal{F}_{-v})$. \footnote{This is straightforward to show  by conditioning $e_{uv}$ on $f_v\in\mathcal{F}_{-v}$, similar to the conditioning in the nominator and denominator of (\ref{eq:lim_cond0}). }

Eqs. (\ref{eq:lim_cond}) and (\ref{eq:lim_cond2}) imply that in the $\frac{K}{N}\rightarrow 0$ regime, the presence of an edge or loop is independent of the presence of the other edges and loops in digraph $\mathcal{D}$ and each is present with probability $q$. This asymptotic model is denoted by $\mathcal{D}_a$.
Given $\mathcal{D}_a$ as the asymptotic model of $\mathcal{D}$ for $\frac{K}{N}\rightarrow 0$, by construction, the undirected graph $\mathcal{G}$ can be asymptotically modeled by $\mathcal{G}_a$ for which every edge is present with probability $q^2$ and every loop is present with probability $q$.

\section{Proof of Theorem~\ref{th:1}}\label{app:mdproof}
To prove Theorem~\ref{th:1}, we use Wormald's theorem following an approach inspired by the approach in \cite[Section~3.4.1]{Mastin:2016}. Let $\mathcal{L}$ be the set of looped vertices of $ \mathcal{G}_a$.  Also, let  $\mathcal{M}$ and $\mathcal{U}$ represent the sets of unlooped vertices that are matched and remain unmatched by Algorithm~\ref{alg:m}, respectively. Since a looped vertex will not be matched by Algorithm~\ref{alg:m}, these three sets are disjoint and partition the vertices of $ \mathcal{G}_a$.

We are interested in the number of looped vertices plus the matchings made by Algorithm~\ref{alg:m}:
\begin{align}\label{eq:rate0}
\frac{|\mathcal{M}|}{2} + |\mathcal{U}| 
= \frac{K-|\mathcal{U}|-|\mathcal{L}|}{2}+|\mathcal{U}|
=\frac{1}{2}(K-|\mathcal{L}|+|\mathcal{U}|).
\end{align} 
This is because one coded message is transmitted per each pair of matched vertices and one uncoded message is transmitted for each unlooped and unmatched vertex. 

Based on (\ref{eq:rate0}), to analyze the statistical behavior of the delivery rate one needs to analyze $|\mathcal{U}|$ and $|\mathcal{L}|$. We do this using Theorem~\ref{th:w} in Appendix~\ref{app:w}. For that, we index the online matching process of Algorithm~\ref{alg:m} by $K$, which is the number of vertices of $ \mathcal{G}_a$. Let us define the two variables $L(t)$ and $U(t)$ on the process.  Variable $L(t)$ is the number of looped vertices in $\{v_1,\cdots,v_{t-1}\}$. Variable $U(t)$ denotes the number of unlooped vertices in $\{v_1,\cdots,v_{t-1}\}$ that are not matched by Algorithm~\ref{alg:m} in the first $t-1$ iterations. Notice that since $U(t)< K$ and $L(t)<K$, we set  $c_0=1$ in Theorem~\ref{th:w}.

In the following, we verify that the three conditions of Theorem~\ref{th:w} are satisfied for the defined variables. Both $L(t)$ and $U(t)$ satisfy the boundedness condition with $\beta=1$ and $\gamma=0$, as $|L(t+1)-L(t)|\leq 1$ and $|U(t+1)-U(t)|\leq 1$ always hold. 
For the trend hypothesis, we have
\begin{align*}
\mathbb{E}(U(t+1)&-U(t)|\mathcal{H}_t) =\\\nonumber
&0\times \mathbb{P}({\small v_t \text{ looped}})\\\nonumber
&-1 \times \mathbb{P}({\small v_t \text{ unlooped and matches at time $\!t$}})\\
&+1 \times \mathbb{P}({\small v_t \text{ unlooped and does not match at time $\!t$}})\\
 & = -(1\!-\!q)(1\!-\!(1\!-\!q^2)^{U(t)}) + (1\!-\!q)(1\!-\!q^2)^{U(t)}\\
 &=(1-q)\left[2(1-q^2)^{U(t)}-1\right].
\end{align*}
and
\begin{align*}
\mathbb{E}(L(t+1)-L(t)|\mathcal{H}_t)=\mathbb{P}(v_t \text{ looped
})=q,
\end{align*}
where $\mathcal{H}_t$ is the history of the process up to time $t$. 
Since the derived expectations are deterministically true, the trend hypothesis holds with $\lambda_1=0$ and $f_1(x,z_1,z_2)=(1-q)[2(1-q^2)^{Kz_1}-1]$ and $f_2(x,z_1,z_2)=q$, with domain $D$ defined as $-\epsilon<x<1+\epsilon$, $-\epsilon<z_1<1+\epsilon$ and $-\epsilon<z_2<1+\epsilon,\,\epsilon>0$. Finally, the Lipschitz hypothesis is satisfied as $f_1$ and $f_2$ are Lipschitz continuous on $\mathbb{R}^2$, because they are differentiable everywhere and have bounded derivatives. 

Since the conditions of the theorem are satisfied, the dynamics of $z_1$ and $z_2$ can be formulated by the differential equations
\begin{align*}
\frac{dz_1(x)}{dx}\!&=\!(1-q)\!\left[2(1-q^2)^{Kz_1(x)}-1\right]\!,\, z_1(0)=0;\\
\frac{dz_2(x)}{dx}&=q,\,z_2(0)=0,
\end{align*}
where the initial conditions result from $U(0)=L(0)=0$. Notice that the equations derived are decoupled in $z_1$ and $z_2$, and can be solved independently as
\begin{align*}
z_1(x)=-\frac{\log(2-(1-q^2)^{K(1-q)x})}{K\log(1-q^2)},\qquad
z_2(x)=qx.
\end{align*}
Then, for $\lambda>\lambda_1+c_0K\gamma=0$, with probability $1-O(\frac{1}{\lambda} e^{-K\lambda^3})$, we have
\begin{align*} 
U(t)= Kz_1(t/K)+O(\lambda K)
=&-\frac{\log(2-(1-q^2)^{(1-q)t})}{\log(1-q^2)}\\&+O(\lambda K),\\
L(t)=Kz_2(t/K)+O(\lambda K)=& qt+O(\lambda K),
\end{align*}
uniformly for $0\leq t \leq K$. 

By evaluating the derived expressions for $U(t)$ and $L(t)$ at $t=K$, and their respective substitution in (\ref{eq:rate0})  for $|\mathcal{U}|$ and $|\mathcal{L}|$, Theorem~\ref{th:1} results.

\section{Proof of Theorem~\ref{th:2}}\label{app:ccproof}
Let $Y_i(t)$ be the number of cliques of size $i$ formed up to iteration $t-1$ when Algorithm~\ref{alg:cc} is applied to the side information graph modeled as $\mathcal{G}_a$,  where $i=1,\ldots,K$. To analyze the behavior of these variables, we use the version of Wormald's theorem that is provided in Remark~\ref{rem:W} in Appendix~\ref{app:w}. This is because the number of variables $Y_i$ depends on $K$ here. 

It is straightforward to show that the boundedness hypothesis of Theorem~\ref{th:w} is satisfied with $\beta=1$ and $\gamma=0$, as in each iteration of the algorithm, the number of cliques of each size either remains unchanged, or decrements or increments by $1$. Furthermore, for the trend hypothesis, we model the expected change of each variable throughout the process as
\begin{align}\nonumber
\mathbb{E}&(Y_i(t+1)-Y_i(t)|\ve{Y}(t)) &\\\nonumber
&=0\times \mathbb{P}({\small v_t \text{ looped}})\\\nonumber
&\quad- \mathbb{P}({\small v_t \text{ unlooped and joins a clique of size $i$ at time $\!t$}})\\\label{eq:trend}
 &\quad + \mathbb{P}({\small v_t \text{ unlooped and joins a clique of size $i-1$ at time $t$}}).
\end{align}
Algorithm~\ref{alg:cc} operates solely based on  the edges and loops of the side information graph. Hence,  the output of the algorithm, i.e., the cliques formed by the algorithm up to time $t$, is a function of the edges and loops of the subgraph over vertices $v_1,\ldots,v_{t-1}$. Since in the asymptotic side information graph $\mathcal{G}_a$ every edge and loop is present independently,  the output of the algorithm up to time $t$ provides no information about the connectivity of vertex $v_t$ to the previously arrived vertices. As a result, $v_t$ is connected to any previously arrived vertex with probability $q^2$.  Hence, with probability $(q^2)^j$, $v_t$ is adjacent to all the vertices in a clique of size $j$. Similarly, $(1-q^{2j})^{Y_j(t)}$ is the probability that none of the $Y_j(t)$ cliques of size $j$  are suitable for $v_t$ at time $t$. Therefore, $\hat{g}_i(\ve{Y}) \delequal \prod_{j=i}^K (1-q^{2j})^{Y_j(t)}$ is the probability that at iteration $t$, there exists no suitable clique of sizes equal or greater than $i$ for $v_t$. Also, $\left(1\!-\!\left(1\!-\!q^{2(i-1)}\right)^{Y_{i-1}(t)}\right)\hat{g}_i(\ve{Y}) $ is the probability that the largest suitable clique for $v_t$ has size $i-1$, which implies that $Y_i(t+1)-Y_i(t)=1$. Finally, since $v_t$ is not looped with probability $(1-q)$, we have $\mathbb{P}({\small v_t \text{ unlooped and joins a clique of size $i-1$ at time $t$}})=(1-q)\left(1-\left(1-q^{2(i-1)}\right)^{Y_{i-1}(t)}\right)\hat{g}_i(\ve{Y})$. Using this result and by following the same line of argument for the case where $Y_i$ decrements at time $t$, (\ref{eq:trend}) simplifies to
 \begin{align}\nonumber
 \mathbb{E}&(Y_i(t+1)-Y_i(t)|\ve{Y}(t)) =(1-q)\times\\\nonumber
 &\left[
-\left(1-\left(1-q^{2i}\right)^{Y_i(t)}\right)\!\prod_{j=i+1}^K \left(1-q^{2j}\right)^{Y_j(t)}\right.\\\label{eq:trend2}
&\;\;+\left.\left(1-\left(1-q^{2(i-1)}\right)^{Y_{i-1}(t)}\right)\prod_{j=i}^K \left(1-q^{2j}\right)^{Y_j(t)}
\right],
\end{align}
where $\ve{Y}(t)=(Y_1(t),\ldots,Y_K(t))$.

Let $f_i(x,z_1,\ldots,z_K)$ be the right hand side of (\ref{eq:trend2}) with $Y_i(t)$ replaced by $Kz_i(x;q)$.\footnote{Notice that  $f_{n-l}$ only depends on $z_n,z_{n-1},z_{n-(l+1)}$. Hence, by an argument similar to the one in Remark~\ref{rem:W}, one can solve for $f_{n-l}$ by restricting the equations to the ones that involve these variables.} Here, we used notation $z_i(\cdot;q)$ to show that every $z_i$ is  parametrized by $q$. Then, the trend hypothesis of Wormald's theorem is satisfied for all variables $Y_i$ and functions $f_i$ with $\lambda_1=0$. 
Functions $f_i$ are differentiable with bounded derivatives, hence they are Lipschitz continuous. Thus, based on Wormald's theorem, we get the system of differential equations in (\ref{eq:diff2}) for $z_i$,
where the initial conditions result from $Y_i(0)=0$. Also, for any $\lambda>0$, we have $Y_i(t)=Kz_i(t/K;q)+O(\lambda K)$ with probability $1-O(\frac{K}{\lambda} e^{-K\lambda^3})$. Hence, (\ref{eq:sys_cc}) results.

\section{Derivation of (\ref{eq:e_r}) and (\ref{eq:e_r_u})}\label{app:exp}
In this appendix, we derive the expressions in (\ref{eq:e_r}) and  (\ref{eq:e_r_u}) that we use to evaluate the expected delivery rate of uncoded caching and the expected rate in \cite[Theorem~2]{Maddah_optimal:2016}. 

As in \cite{Maddah_optimal:2016}, let $N_\text{e}(\ve{d})$ be the number of distinct requests in demand vector $\ve{d}$. Since $\ve{d}$ is a random vector, $N_\text{e}(\ve{d})$ is a random number in $\{1,\ldots,K\}$. Given placement $\mathcal{P}$, if the rate of a delivery algorithm $A_D$ only depends on $N_\text{e}(\ve{d})$ and not the individual files requested in $\ve{d}$, then, the expected delivery rate will be
\begin{align} \label{eq:e_general}
\mathbb{E}_\ve{d}(R_{A_D})=\sum_{m=1}^K  \mathbb{P}(N_\text{e}(\ve{d})=m)R(\ve{d};\mathcal{P},A_D).
\end{align}
Assuming that the popularity distribution of files is uniform, we have
\begin{align} \label{eq:prob}
\mathbb{P}(N_\text{e}(\ve{d})=m)=\frac{\binom{N}{m}\stirling{K}{m}m!}{N^K}.
\end{align}
This is because in $\binom{N}{m}$ ways one can select $m$ out of $N$ files. There are $\stirling{K}{m}$ ways to partition $K$ objects into $m$ non-empty subsets, where $\stirling{K}{m}$ is the Stirling number of the second kind \cite[Section~5.3]{Cameron:1994}. Also, there are $m!$  ways to assign one of the $m$ selected files to each subset. Hence, there are $\binom{N}{m}\stirling{K}{m}m!$ demand vectors  of length $K$ with $m$ distinct files from a library of $N$ files. Since the total number of demand vectors is $N^K$ and they are equiprobable, (\ref{eq:prob}) results.

\paragraph*{Uncoded Caching} Consider the cases of uncoded caching where delivery messages are uncoded. For placement every cache either stores the same set of $M$ files out of the $N$ files in the library, or every cache stores the first $M/NF$ packets of every file. These correspond to uncoded file and subfile cachings. The rate required to delivery demand vector $\ve{d}$ with uncoded messages is $N_\text{e}(\ve{d})(1-M/N)$ files. Hence, based on (\ref{eq:e_general}) and (\ref{eq:prob}), we get
\begin{align}\nonumber
\mathbb{E}_\ve{d}(R_{\text{uncoded}})&=\sum_{m=1}^K  \frac{\binom{N}{m}\stirling{K}{m}m!}{N^K}\times m\left(1-\frac{M}{N}\right)\\\label{eq:ave_uncoded}&
=N\left(1-(1-\frac{1}{N})^K\right)\left(1-\frac{M}{N}\right),
\end{align}
where we used 
\begin{align}\label{eq_Sum1}
\sum_{m=1}^K m\binom{N}{m}\stirling{K}{m}m!=N^{K+1}-N(N-1)^K.
\end{align}
The reason (\ref{eq_Sum1}) holds is as follows.  Consider a set of $K+1$ objects that we want to \textit{partition} into $m$ subsets such that the subset containing the first object has cardinality greater than $1$, i.e., the first object is not the only object in the corresponding subset. This can be done in $m\stirling{K}{m}$ ways as we can partition the rest of the $(K+1)-1$ objects into $m$ subsets and then add the first object to one of the resulting $m$ subsets. 
Assume that the subsets can be labeled distinctly from a set of $N\geq K$ labels. Then, there are $\sum_{m=1}^K m\stirling{K}{m}\binom{N}{m}m!$ ways to partition $K+1$ objects and label them with distinct labels such that the subset containing the first object has cardinality greater than $1$. This sum is equal to the total number of ways to partition $K+1$ objects and distinctly label them with the $N$ available labels, minus the number of ways to do the same but have the first object as the only element in its corresponding subset. The former counts to $N^{K+1}$, and the latter can be done in $N(N-1)^K$ different ways. This proves (\ref{eq_Sum1}).

\paragraph*{Optimal \acrshort{csc}} From (\ref{eq:e_general}), the expected rate in the RHS of \cite[eq.~(27)]{Maddah_optimal:2016} can be written as
\begin{align}\nonumber
\mathbb{E}_\ve{d}(R^*_{\text{CSC}})=\sum_{m=1}^K  &\mathbb{P}(N_\text{e}(\ve{d})=m)\times\\\label{eq:ave_rate_csc}&
\left[\frac{N-M}{M}(1-(1-M/N)^{m})\right],
\end{align}
which by substitution of (\ref{eq:prob}), results in (\ref{eq:e_r}).

\bibliographystyle{ieeetr}
\bibliography{CachingJune16.bib}

\end{document}